\documentclass[prc,aps,float,nofootinbib,superscriptaddress,twocolumn]{revtex4-1}
\usepackage{amssymb}
\usepackage{amsmath}
\usepackage{amsfonts}
\usepackage{epsfig}
\usepackage{bbm}
\usepackage{comment}
\usepackage{multirow}
\usepackage{longtable}
\usepackage{array}
\usepackage{subfigure}
\usepackage{graphics}
\usepackage{array}
\usepackage[usenames,dvipsnames,svgnames,table]{xcolor}
\usepackage{amsxtra}
\usepackage{dsfont} 
\usepackage{enumitem}

\definecolor{bluebb}{RGB}{46,84,206}
\definecolor{redbb}{RGB}{228,30,43}
\definecolor{greenbb}{RGB}{34,139,34}

\newcommand{\bra}[1]{\langle #1 \vert}
\newcommand{\ket}[1]{\vert #1 \rangle}
\newcommand{\scal}[2]{\langle #1 \vert #2 \rangle}
\newcommand{\elma}[3]{\bra{#1} #2 \ket{#3}}

\DeclareMathOperator{\Tr}{Tr}

\allowdisplaybreaks
\begin{document}
\title{On the norm overlap between many-body states. \\ I. Uncorrelated overlap between arbitrary Bogoliubov product states}

\author{B. Bally}
\affiliation{ESNT, IRFU, CEA, Universit\'e Paris - Saclay, F-91191 Gif-sur-Yvette, France}

\author{T. Duguet}
\email{thomas.duguet@cea.fr} 
\affiliation{IRFU, CEA, Universit\'e Paris - Saclay, F-91191 Gif-sur-Yvette, France}
\affiliation{KU Leuven, Instituut voor Kern- en Stralingsfysica, 3001 Leuven, Belgium}
\affiliation{National Superconducting Cyclotron Laboratory and Department of Physics and Astronomy, Michigan State University, East Lansing, MI 48824, USA}

\date{\today}

\begin{abstract} 
\begin{description}
\item[Background] State-of-the-art multi-reference energy density functional (MR-EDF) calculations require the computation of norm overlaps between different Bogoliubov quasiparticle many-body states. It is only recently that the efficient and unambiguous calculation of such norm kernels has become available under the form of Pfaffians~[L. M. Robledo, Phys. Rev. C79, 021302 (2009)]. Recently developped particle-number-restored Bogoliubov coupled-cluster (PNR-BCC) and particle-number-restored many-body perturbation (PNR-BMBPT) ab initio theories~[T. Duguet and A. Signoracci, J. Phys. G44, 015103
(2017)] make use of generalized norm kernels incorporating explicit many-body correlations. In PNR-BCC and PNR-BMBPT, the Bogoliubov states involved in the norm kernels differ specifically via a global gauge rotation. 
\item[Purpose] The goal of this work is three-fold. We wish (i) to propose and implement an alternative to the Pfaffian method to compute unambiguously the norm overlap between {\it arbitrary} Bogoliubov quasiparticle states, (ii) to extend the first point to {\it explicitly correlated} norm kernels, and (iii) to scrutinize the analytical content of the correlated norm kernels employed in PNR-BMBPT. Point (i) constitutes the purpose of the present paper while points (ii) and (iii) are addressed in a forthcoming companion paper. 
\item[Methods] We generalize the method used in~[T. Duguet and A. Signoracci, J. Phys. G44, 015103
(2017)] in such a way that it is applicable to kernels involving arbitrary pairs of Bogoliubov states. The formalism is presently explicited in detail in the case of the {\it uncorrelated} overlap between arbitrary Bogoliubov states. The power of the method is numerically illustrated and benchmarked against known results on the basis of toy models of increasing complexity. 
\item[Results] The norm overlap between arbitrary Bogoliubov product states is obtained under a closed-form expression allowing its computation without any phase ambiguity. The formula is physically intuitive, accurate and versatile. It equally applies to norm overlaps between Bogoliubov states of even or odd number parity. Numerical applications illustrate these features and provide a transparent representation of the content of the norm overlaps. 
\item[Conclusions] The complex norm overlap between arbitrary Bogoliubov states is computed, without any phase ambiguity, via elementary linear algebra operations. The method can be used in any configuration mixing of orthogonal and non-orthogonal product states. Furthermore, the closed-form expression extends naturally to correlated overlaps at play in PNR-BCC and PNR-BMBPT. As such, the straight overlap between Bogoliubov states is the zeroth-order reduction of more involved norm kernels to be studied in a forthcoming paper.

\end{description}
\end{abstract}
\maketitle

%

\section{Introduction}
%
%

In the vast majority of methods, the A-body Schr\"{o}dinger equation is (approximately) solved by representing it on a (truncated) orthonormal basis of the A-body Hilbert space ${\cal H}_A$. Some approaches, however, represent the A-body Schr\"{o}dinger equation on a finite-dimensional set of non-orthogonal states of ${\cal H}_A$. The set in question may even exceed ${\cal H}_A$ by employing states mixing vectors belonging to Hilbert spaces associated with different particle numbers, i.e. states that are genuine vectors of Fock space ${\cal F}$. This is for instance the case of the generator coordinate method with/without symmetry restoration that underlines state-of-the-art multi-reference energy density functional (MR-EDF) calculations~\cite{ring80a,bender03b,Duguet:2013dga,Egido:2016bdz}. In this method, an (effective) Hamilton operator is diagonalized within a manifold of non-orthogonal Bogoliubov product states such that the secular equation to be solved requires the evaluation of the norm matrix constructed from overlaps between all members of the manifold.

The evaluation of the overlap between two non-orthogonal product states has a long history. The overlap between two non-orthogonal A-body Slater determinants poses no problem and has long been known to be computable as a determinant~\cite{blaizot86}. Contrarily, it is only recently that the efficient and unambiguous calculation of the overlap between two arbitrary Bogoliubov quasiparticle states has become available as the Pfaffian of a skew-symmetric matrix~\cite{Robledo:2009yd,Robledo:2011ce,Avez:2011wr}. As of right now, this is the most general and efficient method available to MR-EDF practitioners.

It so happens that the recently developped particle-number-restored Bogoliubov coupled-cluster (PNR-BCC) and particle-number-restored many-body perturbation (PNR-BMBPT) theories~\cite{Duguet:2015yle} also build on a manifold of non-orthogonal Bogoliubov states. These formalisms involve norm kernels that are more general that the straight overlap between the non-orthogonal Bogoliubov states at play. The norm kernels explicitly incorporate many-body correlations and reduce to the mere overlap between two non-orthogonal Bogoliubov states whenever such correlations are omitted. These more general norm kernels are thus presently denoted as {\it correlated} norm kernels whereas straight overlaps between Bogoliubov vacua are characterized as {\it uncorrelated} norm kernels.

In this context, our objective is to derive and test a closed-form expression of norm kernels (i) providing an alternative to the Pfaffian method for the overlap between arbitrary Bogoliubov quasiparticle states and (ii) naturally extending to the computation of correlated norm kernels. After deriving the closed-form formula for generic norm kernels, the present paper focuses on its analytical explicitation and on its numerical implementation in the particular case of uncorrelated kernels. The analytical examination and the numerical testing of correlated norm kernels appearing within the frame of PNR-BMBPT are left to a forthcoming companion paper~\cite{arthuis17a}. 

The present paper is organized as follows. In Sec.~\ref{generalformalism}, the closed-form formula is derived by generalizing the method used in Ref.~\cite{Duguet:2015yle}. Section~\ref{formalismkernel} is dedicated to the formal implementation of the master formula in the particular case of uncorrelated kernels. Following the formal set up, Sec.~\ref{results} proposes numerical illustrations of the validity and the versatility of the presently-proposed formula on the basis of toy models of increasing complexity. Section~\ref{manybodymethods} discusses more specifically how the calculation of uncorrelated norm kernels enters typical generator coordinate method and symmetry restoration calculations. Eventually, conclusions are given in Sec.~\ref{Conclusions}. Three appendices provide relevant technical details.

\section{General formalism}
\label{generalformalism}

\subsection{Many-body problem}

Let us consider a A-body fermionic system governed by the hamiltonian $H$ to which is associated the grand potential $\Omega = H -\lambda A$, where $\lambda$ is the chemical potential and $A$ the particle-number operator. Energy eigenstates are solution of 
\begin{subequations}
\label{equationschroedinger}
\begin{align}
H | \Psi^{\text{A}}_k \rangle &= \text{E}^{\text{A}}_k | \Psi^{\text{A}}_k \rangle \, , \\
\Omega | \Psi^{\text{A}}_k \rangle &= \Omega^{\text{A}}_k | \Psi^{\text{A}}_k \rangle \, ,
\end{align}
\end{subequations}
with $\Omega^{\text{A}}_k = \text{E}^{\text{A}}_k - \lambda  \text{A}$. Given a set of single-particle creation and annihilation operators $\left\{ c_k ; c^{\dagger}_k \right\}$ associated with an arbitrary basis of the one-body Hilbert space ${\cal H}_1$, the Hamiltonian\footnote{The formalism can be extended to a Hamiltonian containing three- and higher-body forces without running into any fundamental problem. Also, one subtracts the center of mass kinetic energy to the Hamiltonian in actual calculations of finite nuclei. As far as the present work is concerned, this simply leads to a redefinition of one- and two-body matrix elements $t_{pq}$ and $\bar{v}_{pqrs}$ in the Hamiltonian without changing any aspect of the many-body formalism that follows.} reads as
\begin{equation}
H \equiv  \frac{1}{(1!)^2} \sum _{pq} t_{pq} c^{\dagger}_{p} c_{q} + \frac{1}{(2!)^2} \sum _{pqrs} \bar{v}_{pqrs}  c^{+}_{p} c^{+}_{q} c_{s} c_{r}  \ , \label{e:ham} 
\end{equation} 
where antisymmetric matrix elements of the two-body interaction are implied.

\subsection{Multi-reference set}

The situation of typical interest relies on the use of a manifold of $N_{\text{set}}$ Bogoliubov product states of identical\footnote{The approach designed below can be extended to the mixing of states carrying different number parity at the price of considering improper Bogoliubov transformations. For simplicity, we do not consider this case in the present work.} number parity~\cite{ring80a}
\begin{equation}
\mathcal{M} \equiv \big\{ \ket{\Phi_{1}}, \ldots, \ket{\Phi_{N_{\text{set}}}}\big\} \, ,
\end{equation}
employed in a method designed to solve Eq.~\ref{equationschroedinger}.  Below, we employ two generic states $| \Phi \rangle$ and $| \breve{\Phi} \rangle$ to denote {\it any} pair of states belonging to $\mathcal{M}$. 

\subsection{Purpose of the study}

Our most general goal is to compute the norm kernel
\begin{equation}
{\cal N}(\tau) \equiv \frac{\langle \Psi(\tau)| \breve{\Phi} \rangle }{\langle \Psi(\tau)| \Phi \rangle} \, , \label{defgeneralnorm}
\end{equation}
where 
\begin{subequations}
\begin{align}
| \Psi(\tau) \rangle & \equiv e^{-\tau \Omega} |  \Phi \rangle \, \label{defpsi} \\
| \breve{\Phi} \rangle  & \equiv e^{iS} |  \Phi \rangle \, , \label{defphibreve}
\end{align}
\end{subequations}
with $S$ a general one-body hermitian operator acting on Fock space
\begin{align}
 S &\equiv s^{00} + s^{11} + s^{20} + s^{02} \label{eq:1bodynopart} \\
   &= s^{00} + \sum_{pq}  s^{11}_{pq} c^{\dagger}_{p} c_{q} + \frac12 \sum_{pq}  \left\{ s^{20}_{pq} c^{\dagger}_{p} c^{\dagger}_{q} 
                                                                                              + s^{02}_{pq} c_{q} c_{p} \right\} \nonumber \\
&= s^{00} + \frac12 \Tr\left(s^{11}\right) 
+ \frac12  
\left(\,c^{\dagger} \hspace{0.2cm} c \,\right)
\begin{pmatrix}
s^{11} & s^{20} \\
-s^{02} &  -s^{11 \ast}
\end{pmatrix} 
\begin{pmatrix}
c \\
c^{\dagger}
\end{pmatrix} \, , \nonumber
\end{align}
where annihilation and creation operators are organized in vectors of twice the dimension $N$ of ${\cal H}_1$. In Eq.~\ref{eq:1bodynopart}, $s^{00}$ is a real number, $s^{11}$ is a hermitian matrix whereas $s^{20}$ and $s^{02}$ are skew-symmetric matrices satisfying \mbox{$s^{02}=s^{20 \ast}$}. These conditions make the matrix
\begin{equation}
s \equiv \begin{pmatrix}
s^{11} & s^{20} \\
-s^{02} &  -s^{11 \ast}
\end{pmatrix} 
\end{equation}
to be hermitian. 

Strictly speaking, ${\cal N}(\tau)$ provides the {\it ratio} of the two overlaps computed between state $| \Psi(\tau) \rangle$ and the Bogoliubov states $| \Phi \rangle$ and $| \breve{\Phi} \rangle$. Presently, we are interested in the case where $| \Psi(\tau) \rangle$ is obtained via the imaginary-time propagation of $| \Phi \rangle$ over the interval $[0,\tau]$; i.e. where $| \Psi(\tau) \rangle$ is expanded around the unperturbed reference state $| \Phi \rangle$. The overlap $\langle \Psi(\tau)| \Phi \rangle$ appearing in the denominator of Eq.~\ref{defgeneralnorm} is said to possess a {\it diagonal} character in the sense that the same state $| \Phi \rangle$ is used both in the bra and in the ket. Standard many-body methods are implemented on the basis of such diagonal kernels for which {\it intermediate normalization} is typically used, i.e. the diagonal overlap is set to one. Alternatively, $\langle \Psi(\tau)| \Phi \rangle$ can be calculated via standard many-body techniques\footnote{Either way, the diagonal overlap $\langle \Psi(\tau)| \Phi \rangle$ reduces to $\langle \Phi | \Phi \rangle =1$ when many-body correlations are omitted, i.e. when setting $\tau = 0$.}, e.g. MBPT~\cite{blaizot86}. The non-trivial character of ${\cal N}(\tau)$ relates to its numerator that uses a different Bogoliubov state in the ket from the one used to expand $\langle \Psi(\tau) |$. As long as $| \Phi \rangle$ and $| \breve{\Phi} \rangle$ differ non-trivially, the numerator, and thus ${\cal N}(\tau)$ itself, is said to possess an {\it off-diagonal} character. Eventually, ${\cal N}(\tau)$ is meant to deliver the off-diagonal overlap $\langle \Psi(\tau)| \breve{\Phi} \rangle$  {\it relative} to the known, i.e. already accessible via standard techniques, diagonal overlap $\langle \Psi(\tau)| \Phi \rangle$.

Expressing $| \breve{\Phi} \rangle$ in terms of $| \Phi \rangle$,  Eq.~\ref{defphibreve} specifies the nature of the unitary transformation linking both Bogoliubov states. Accordingly, $| \breve{\Phi} \rangle$ is said to be the {\it transformed} reference state. As demonstrated later on, the {\it proper} transformation over Fock space parameterized by a general hermitian one-body operator $S$ conserving number parity does qualify as a way to represent the unitary connection between two arbitrary Bogoliubov states of identical number parity. The method designed to extract $S$ in the following generalizes previous attempts~\cite{ring77a,hara79a}.

Last but not least, let us remark that the time propagation driven by the interacting Hamiltonian $H$ forbids $| \Psi(\tau) \rangle$ to retain the simplicity of a Bogoliubov product state, which characterizes the {\it correlated} nature of the norm kernel ${\cal N}(\tau)$.

\subsection{Cases of interest}
\label{discussion2}

We wish to compute the norm kernel ${\cal N}(\tau)$ in two particular limits
\begin{enumerate}
\item At $\tau = 0$, the norm kernel reduces to the {\it uncorrelated} off-diagonal overlap
\begin{equation}
{\cal N}(0) = \frac{\langle \Phi | \breve{\Phi} \rangle }{\langle \Phi | \Phi \rangle} 
\end{equation}
between the reference Bogoliubov state and its transformed partner. This is the case we focus on in the present paper.
\item At $\tau = \infty$, the norm kernel involves the exact many-body ground-state\footnote{The chemical potential $\lambda$ is fixed such that $\Omega^{\text{A}_0}_0$ for the targeted particle number $\text{A}_0$ is the lowest value of all $\Omega^{\text{A}}_k$ over Fock space, i.e. it penalizes systems with larger number of particles such that $\Omega^{\text{A}_0}_0 < \Omega^{\text{A}}_\mu$ for all $\text{A}>\text{A}_0$ while maintaining at the same time that $\Omega^{\text{A}_0}_0 < \Omega^{\text{A}}_\mu$ for all $\text{A}<\text{A}_0$. This is achievable if $\text{E}^{\text{A}}_0$ is strictly convex in the neighborhood of $\text{A}_0$, which is generally but not always true for atomic nuclei.}
\begin{equation}
{\cal N}(\infty) = \frac{\langle \Psi^{\text{A}}_0 | \breve{\Phi} \rangle }{\langle \Psi^{\text{A}}_0 | \Phi \rangle} \, .
\end{equation}
A forthcoming companion paper~\cite{arthuis17a} is dedicated to the evaluation and the analysis of correlated off-diagonal kernels whenever correlations are evaluated within many-body perturbation theory~\cite{Duguet:2015yle}.
\end{enumerate}

\subsection{Master formula}
\label{master}

Given $| \Phi \rangle$ and $| \breve{\Phi} \rangle$ belonging to $\mathcal{M}$, the strategy to obtain the desired norm kernel relies on building an auxiliary manifold according to
\begin{equation}
\mathcal{M}[| \Phi \rangle,S] \equiv \big\{ \ket{\Phi(\theta)} \equiv e^{i \theta S}  \ket{\Phi} \, , \, \theta \in [0, 1] \big\} \, ,
\end{equation}
and containing $\ket{\Phi}$ (resp. $| \breve{\Phi} \rangle$) at its origin (resp. end) given that $\ket{\Phi(0)}=\ket{\Phi}$ (resp. $\ket{\Phi(1)}= | \breve{\Phi} \rangle$). The transformation $e^{i \theta S}$ being a one-body unitary transformation, all states in $\mathcal{M}[| \Phi \rangle,S]$ are Bogoliubov vacua~\cite{blaizot86} as will be illustrated later on.

Next, a norm kernel defined along the auxiliary manifold is introduced, for an {\it arbitrary} many-body state $| \Theta \rangle$, via
\begin{equation}
{\cal N}[\langle \Theta |,| \Phi(\theta) \rangle] \equiv \frac{\langle \Theta | \Phi(\theta) \rangle }{\langle \Theta | \Phi \rangle} \, ,
\end{equation}
such that ${\cal N}[\langle \Theta |,| \Phi(0) \rangle]=1$. Differentiating with respect to $\theta$ leads to
\begin{equation}
\label{intermediate1}
\begin{split}
 \frac{d}{d\theta} \, {\cal N}[\langle \Theta |,| \Phi(\theta) \rangle] &= \frac{\elma{\Theta}{i S e^{i \theta S}}{\Phi}}{\langle \Theta | \Phi \rangle} \\
                                               &= i \frac{\elma{\Theta}{S}{\Phi(\theta)}}{\langle \Theta | \Phi \rangle} \, .
\end{split}
\end{equation}
Assuming that ${\cal N}[\langle \Theta |,| \Phi(\theta) \rangle] \ne 0$ for $\theta \in [0, 1]$, one divides both sides of Eq.~\ref{intermediate1} by it to obtain 
\begin{equation}
 \frac{d}{d\theta}\ln \Big({\cal N}[\langle \Theta |,| \Phi(\theta) \rangle]\Big) = i s[\langle \Theta |,| \Phi(\theta) \rangle] \, ,
\end{equation}
with the short-hand notation 
\begin{equation}
s[\langle \Theta |,| \Phi(\theta) \rangle] \equiv  \frac{\elma{\Theta}{S}{\Phi(\theta)}}{\scal{\Theta}{\Phi (\theta)}} \, .
\end{equation}
Integrating the above first-order differential equation between $0$ and $\theta$ leads to
\begin{equation}
 \Big[ \ln \Big({\cal N}[\langle \Theta |,| \Phi(\phi) \rangle] \Big) \Big]^\theta_0 = i \int_0^\theta d\phi \,  \frac{\elma{\Theta}{S}{\Phi(\phi)}}{\scal{\Theta}{\Phi (\phi)}} \, , 
\end{equation}
which rewrites as
\begin{equation}
\label{eq:genov}
{\cal N}[\langle \Theta |,| \Phi(\theta) \rangle] = e^{i\int_0^\theta d\phi \, s[\langle \Theta |,| \Phi(\phi) \rangle]} \, .
\end{equation}
Equation~\ref{eq:genov} constitutes the master formula repeatedly used throughout the present work. 

\subsection{Norm kernels}
\label{resultkernel}

Setting $\langle \Theta |\equiv\langle \Psi(\tau) |$ and $\theta=1$ in Eq.~\ref{eq:genov}, the correlated norm kernel at time $\tau$ is obtained as
\begin{equation}
\frac{\langle \Psi(\tau)| \breve{\Phi} \rangle }{\langle \Psi(\tau)| \Phi \rangle}  =  e^{i\int_0^1 d\phi \, s[\langle \Psi(\tau) |,| \Phi(\phi) \rangle]} \, , \label{eq:genov2}
\end{equation}
where 
\begin{equation}
s[\langle \Psi(\tau) |,| \Phi(\theta) \rangle] =  \frac{\elma{\Psi(\tau)}{S}{\Phi(\theta)}}{\scal{\Psi(\tau)}{\Phi (\theta)}} \, ,
\end{equation}
denotes the so-called {\it linked-connected}~\cite{Duguet:2014jja,Duguet:2015yle}  kernel of the operator $S$ along the manifold $\mathcal{M}[| \Phi \rangle,S]$ at time $\tau$. We note that $s[\langle \Psi(\tau) |,| \Phi(\theta) \rangle]$ is independent of the relative phase between $| \Phi \rangle$ and $| \Phi(\theta) \rangle$.

Employing $\langle \Theta |\equiv \langle \Psi^{\text{A}}_0 |$ and $\theta=1$ in Eq.~\ref{eq:genov}, or equivalently setting $\tau=+\infty$ in Eq.~\ref{eq:genov2}, the ratio of the overlaps between the correlated ground-state and the two Bogoliubov vacua $| \Phi \rangle$ and $| \breve{\Phi} \rangle$ is given by 
\begin{equation}
\frac{\langle \Psi^{\text{A}}_0| \breve{\Phi} \rangle }{\langle \Psi^{\text{A}}_0| \Phi \rangle}  =  e^{i\int_0^1 d\phi \, s[\langle \Psi^{\text{A}}_0 |,| \Phi(\phi) \rangle]} \, , \label{eq:genov2GS}
\end{equation}
where 
\begin{equation}
s[\langle \Psi^{\text{A}}_0 |,| \Phi(\theta) \rangle] =  \frac{\elma{\Psi^{\text{A}}_0}{S}{\Phi(\theta)}}{\scal{\Psi^{\text{A}}_0}{\Phi (\theta)}} \, ,
\end{equation}
denotes the {\it ground-state} linked-connected kernel of the operator $S$ along the manifold $\mathcal{M}[| \Phi \rangle,S]$.

Omitting many-body correlations, i.e. employing $\langle \Theta |\equiv\langle \Phi |$ and $\theta=1$ in Eq.~\ref{eq:genov} or equivalently setting $\tau=0$ in Eq.~\ref{eq:genov2}, the uncorrelated norm kernel between two arbitrary Bogoliubov vacua is written as
\begin{equation}
\label{eq:genovHFB}
\frac{\langle \Phi | \breve{\Phi} \rangle}{\langle \Phi | \Phi \rangle}  = e^{i\int_0^1 d\phi \, s[\langle \Phi |,| \Phi(\phi) \rangle] } \, ,
\end{equation}
where the {\it uncorrelated} linked-connected\footnote{The uncorrelated off-diagonal kernel of an operator is trivially linked and connected~\cite{Duguet:2014jja,Duguet:2015yle}. One can thus omit this qualifier when dealing with such an uncorrelated operator kernel.} kernel of the operator $S$ obviously reads as
\begin{equation}
\label{eq:genovHFBbis}
s[\langle \Phi |,| \Phi(\theta) \rangle] =  \frac{\elma{\Phi}{S}{\Phi(\theta)}}{\scal{\Phi}{\Phi (\theta)}} \, .
\end{equation}
The simplest form given in Eqs.~\ref{eq:genovHFB}-\ref{eq:genovHFBbis} already allows one to address many situations of interest, i.e. it accesses the norm overlaps encountered in symmetry-projected Hartree-Fock-Bogoliubov calculations and/or in generator coordinate calculations along an arbitrary collective coordinate~\cite{bender03b,Duguet:2013dga,Egido:2016bdz}. 

Correlated and uncorrelated norm kernels, including their complex phase, have thus been powerfully expressed in a simple and compact form via the integration of the off-diagonal kernel of the operator $S$ along the auxiliary manifold linking $| \Phi \rangle$ to $| \breve{\Phi} \rangle$. The rest of the present paper is dedicated to the formal and numerical implementations of Eq.~\ref{eq:genovHFB}, with the Bogoliubov transformations defining $| \Phi \rangle$ and $| \breve{\Phi} \rangle$ as the sole inputs. The procedure consists of extracting the operator $S$ and computing its off-diagonal kernel along the auxiliary manifold $\mathcal{M}[| \Phi \rangle,S]$ before proceeding to the integration in Eq.~\ref{eq:genovHFB}.

\subsection{Phase convention}
\label{discussion1}

The normalized states belonging to $\mathcal{M}$ are all individually defined up to a phase. In order to compute consistently the $N_{\text{set}}(N_{\text{set}}+1)/2$ norm kernels within the set, one must fix their relative phases in a synchronized fashion. This can be done by specifying the phase each member of the set shares with a common known state of reference generically denoted as $| \bar{\Phi} \rangle$. There is an entire freedom in choosing the values of those $N_{\text{set}}$ reference phases since no physical observable can depend on them. In order for the latter to be true, the $N_{\text{set}}(N_{\text{set}}+1)/2$ norm kernels must however be consistently computed on the basis of this conventional choice. 

Although it is not mandatory to do so, we presently please to choose the $N_{\text{set}}$ phases relative to the common reference state $| \bar{\Phi} \rangle$ to be identical, i.e. we will require that
\begin{equation}
\text{Arg} (\langle \bar{\Phi} | \Phi_{1} \rangle) = \text{Arg} (\langle \bar{\Phi} | \Phi_{2} \rangle) \ldots = \text{Arg} (\langle \bar{\Phi} | \Phi_{N_{\text{set}}} \rangle) \, . \label{phasecond1}
\end{equation}
Before translating Eq.~\ref{phasecond1} into a mathematical constraint, let us make four comments regarding its practical implications
\begin{itemize}
\item Condition~\ref{phasecond1} only stipulates that all states entertain the same phase relative to a given reference state without fixing/determining the {\it value} of this common phase. This is sufficient to perform the consistent and unambiguous calculation of any physical osbervable.
\item Practically speaking, the above requirement is consistent with the fact that the many-body states involved are presently expressed  {\it relative} to one another via the unitary transformations that connect them (see Eq.~\ref{defphibreve}). This makes unnecessary to explicitly represent the states themselves, which would fix their phases in an absolute sense. In particular, this has the subtile consequence that multiplying one of the states involved with an arbitrary phase implicitly propagates to all the other states involved, thus leaving the norm kernels untouched.
\item When benchmarking the present approach against known results, e.g. employing Pfaffians~\cite{Robledo:2009yd,Robledo:2011ce,Avez:2011wr}, one will have to pay particular attention to the fact that these methods usually rely on an explicit representation of the two states involved that do not necessarily match the condition elaborated above. In particular, multiplying one of the two states with an arbitrary phase typically induces a change of the norm overlap by the same phase, which is not the case here. We will elaborate further on this point in Sec.~\ref{family}.
\item In practice, an appropriate reference state has to be specified. A natural choice is to choose the particle vacuum, e.g. to set $| \bar{\Phi} \rangle \equiv | 0 \rangle$. This choice is often convenient but is not appropriate whenever the states making up  $\mathcal{M}$ are intrinsically orthogonal to $|0 \rangle$, e.g. for Bogoliubov states used to describe systems constituted by an odd number of fermions. A generically suitable strategy is to use a state within $\mathcal{M}$ as the state of reference, e.g. to set $| \bar{\Phi} \rangle \equiv | \Phi_{1} \rangle$. In this way, the reference state naturally shares basic symmetries carried by the Bogoliubov states making up $\mathcal{M}$ and is less likely to be orthogonal to them, i.e. it is likely to be only accidentally orthogonal to some of them. These two choices will be illustrated both formally and numerically in the following.
\end{itemize}

Focusing on an arbitrary pair $| \breve{\Phi} \rangle$ and $| \Phi \rangle$ belonging to $\mathcal{M}$, Eq.~\ref{phasecond1} can be translated into a more explicit mathematical condition. Exploiting Eq.~\ref{eq:genov} for $\langle \Theta | \equiv \langle \bar{\Phi} |$ and $\theta = 1$, one obtains
\begin{equation}
\frac{\langle \bar{\Phi} | \breve{\Phi} \rangle}{\langle \bar{\Phi} | \Phi \rangle} = e^{-\Im m \int_0^1 d\phi \, s[\langle \bar{\Phi} |,| \Phi(\phi) \rangle]} \, e^{i \Re e\int_0^1 d\phi \, s[\langle \bar{\Phi} |,| \Phi(\phi) \rangle]}  \, ,
\end{equation}
such that the requirement that $| \Phi \rangle$ and $| \breve{\Phi} \rangle$ entertain the same phase with $| \bar{\Phi} \rangle$  rewrites as
\begin{equation}
\Re e\int_0^1 d\theta \, \, s[\langle \bar{\Phi} |,| \Phi(\theta) \rangle] =0 \, . \label{phasecondition1}
\end{equation}
In view of this condition, let us distinguish the two practical cases of interest.
\begin{enumerate}
\item  The operator $S$ relating $| \breve{\Phi} \rangle$ to $| \Phi \rangle$ is a given of the problem, e.g. $| \breve{\Phi} \rangle$ is explicitly obtained from $|  \Phi \rangle$ via a one-body symmetry transformation whose unitary representation in terms of $S$ is known from prior considerations. In this situation, it must be checked that the operator $S$ is such that Eq.~\ref{phasecondition1} is fulfilled. If this is the case\footnote{With $S$ defining a one-body symmetry transformation and the choice $| \bar{\Phi} \rangle \equiv | 0 \rangle$, the identity $\langle 0 | \breve{\Phi} \rangle = \langle 0 | \Phi \rangle$ is automatically satisfied, implying the validity of the phase convention.}, nothing more is to be considered. If this is not the case, one must enforce Eq.~\ref{phasecondition1} by multiplying the unitary transformation linking $| \breve{\Phi} \rangle$ to $| \Phi \rangle$ (Eq.~\ref{defphibreve}) by the corresponding phase difference. See Sec.~\ref{symmetryrestoration} for a typical example.
\item The Bogoliubov transformations (see Sec.~\ref{bogoalbegra} below) defining $| \Phi \rangle$ and $| \breve{\Phi} \rangle$ are the given of the problem. In this situation, the operator $S$ relating both states is to be extracted. As explained later on, this can only be done up to the arbitrary constant $s^{00}$ that is to be fixed thanks to Eq.~\ref{phasecondition1}.
\end{enumerate}

\section{Uncorrelated norm kernel}
\label{formalismkernel}

For the remaining of the present paper, the focus is on the uncorrelated norm kernel given by Eqs.~\ref{eq:genovHFB},~\ref{eq:genovHFBbis} and~\ref{phasecondition1}. In practice, the evaluation of the norm kernel makes necessary to express these three equations in terms of the inputs to the problem, i.e. the ingredients characterizing the two Bogoliubov states $\ket{\Phi}$ and  $\ket{\breve{\Phi}}$. 

\subsection{Bogoliubov transformations}
\label{bogoalbegra}

Typically, $\ket{\Phi}$ and  $\ket{\breve{\Phi}}$ are introduced as vacua of two sets of quasi-particle operators $\left\{ \beta_k ; \beta^{\dagger}_k \right\}$ and $\left\{ \breve{\beta}_k ; \breve{\beta}^{\dagger}_k \right\}$ defined through unitary linear Bogoliubov transformations\footnote{The present work deals with so-called {\it proper} Bogoliubov transformations connecting states with the same number parity. The matrices of proper Bogoliubov transformation form an isomorphic group to the group $SO(2N)$ of orthogonal matrices with determinant +1~\cite{blaizot86}.} 
\begin{subequations}
\begin{align}
\begin{pmatrix}
\beta \\
\beta^{\dagger}
\end{pmatrix}
&= {\cal W}^{\dagger} 
\begin{pmatrix}
c \\
c^{\dagger}
\end{pmatrix}  \, , \label{bogophi} \\
\begin{pmatrix}
\breve{\beta} \\
\breve{\beta}^{\dagger}
\end{pmatrix}
&= \breve{{\cal W}}^{\dagger} 
\begin{pmatrix}
c \\
c^{\dagger}
\end{pmatrix}  \, ,
\end{align}
\end{subequations}
where
\begin{subequations}
\begin{align}
{\cal W} &\equiv
\begin{pmatrix}
U & V^{\ast} \\
V &  U^{\ast}
\end{pmatrix}  \, , \\
\breve{{\cal W}} &\equiv
\begin{pmatrix}
\breve{U} & \breve{V}^{\ast} \\
\breve{V} &  \breve{U}^{\ast}
\end{pmatrix} \,  . \label{bogomatrix}
\end{align}
\end{subequations}

The expanded form of Eq.~\ref{bogophi} reads as
\begin{subequations}
\begin{align}
 \beta_{k_1} &= \sum_{k_2} U^*_{k_2 k_1} \, c_{k_2} + V^*_{k_2 k_1} \, c_{k_2}^\dagger \, ,  \\
 \beta_{k_1}^\dagger &= \sum_{k_2} U_{k_2 k_1} \, c_{k_2}^\dagger + V_{k_2 k_1} \, c_{k_2} \, ,  
\end{align}
\end{subequations}
and similarly for $\left\{  \breve{\beta}_k ; \breve{\beta}^{\dagger}_k\right\}$. 

The unitarity of ${\cal W}$ ensures that quasi-particle operators $\left\{  \beta_k ; \beta^{\dagger}_k\right\}$ fulfill fermionic anti-commutation rules, which leads to four relations 
\begin{subequations}
\label{unitarity}
\begin{align}
UU^{\dagger} + V^{\ast}V^{T} &= 1 \ , \label{unitarityA} \\
VU^{\dagger} + U^{\ast}V^{T} &= 0 \ , \label{unitarityB} \\
U^{\dagger}U + V^{\dagger}V  &= 1 \ , \label{unitarityC} \\
V^{T}U + U^{T}V  &= 0 \, , \label{unitarityD} 
\end{align}
originating from ${\cal W}^{\dagger}{\cal W} = 1$ and to four relations
\begin{align}
UV^{\dagger} + V^{\ast}U^{T} &= 0 \ , \label{unitarityE} \\
VV^{\dagger} + U^{\ast}U^{T} &= 1 \ , \label{unitarityF} \\
U^{\dagger}V^{\ast} + V^{\dagger}U^{\ast}  &= 0 \ , \label{unitarityG} \\
V^{T}V^{\ast} + U^{T}U^{\ast}  &= 1 \ , \label{unitarityH} 
\end{align}
\end{subequations}
originating from ${\cal W}{\cal W}^{\dagger} = 1$. Similar relationships hold for $\breve{U}$ and $\breve{V}$.

One further introduces the skew-symmetric matrices
\begin{subequations}
\begin{align}
Z \equiv V^{\ast}[U^{\ast}]^{-1} \, , \\
\breve{Z} \equiv \breve{V}^{\ast}[\breve{U}^{\ast}]^{-1} \, ,
\end{align}
\end{subequations}
in terms of which $\ket{\Phi}$ and $\ket{\breve{\Phi}}$ can be expressed with respect to $| 0 \rangle$ by virtue of Thouless' theorem~\cite{thouless60}. 

\subsection{Representing $S$ in the quasi-particle basis of $\ket{\Phi}$}
\label{sec:onebody}

For later use, the one-body operator $S$ is reexpressed in the quasi-particle basis $\left\{ \beta_k ; \beta^{\dagger}_k \right\}$ associated with $\ket{\Phi}$
\begin{align}
 S &\equiv S^{00} + S^{11} + S^{20} + S^{02} \label{eq:1bodynobogo} \\
   &= S^{00} + \sum_{k_1 k_2}  S^{11}_{k_1 k_2} \beta^{\dagger}_{k_1} \beta_{k_2} \nonumber \\
   & \,\,\,\,\,\, + \frac12 \sum_{k_1 k_2}  \left\{ S^{20}_{k_1 k_2} \beta^{\dagger}_{k_1} \beta^{\dagger}_{k_2} 
                                                                                              + S^{02}_{k_1 k_2} \beta_{k_2} \beta_{k_1} \right\} \nonumber \\
&= S^{00} + \frac12 \Tr\left(S^{11}\right) 
+ \frac12  
\left(\,\beta^{\dagger} \hspace{0.2cm} \beta \,\right)
\begin{pmatrix}
S^{11} & S^{20} \\
-S^{02} &  -S^{11 \ast}
\end{pmatrix} 
\begin{pmatrix}
\beta \\
\beta^{\dagger}
\end{pmatrix}  \, , \nonumber
\end{align}
where $S^{00}$ is a real number, $S^{11}$ is a hermitian matrix whereas $S^{20}$ and $S^{02}$ are skew-symmetric matrices satisfying \mbox{$S^{02}=S^{20 \ast}$}. These conditions make the matrix
\begin{equation}
{\cal S} \equiv \begin{pmatrix}
S^{11} & S^{20} \\
-S^{02} &  -S^{11 \ast}
\end{pmatrix} 
\end{equation}
to be hermitian. Matrices appearing in the single-particle (Eq.~\ref{eq:1bodynopart}) and the quasi-particle (Eq.~\ref{eq:1bodynobogo}) representations of $S$ are related via
\begin{subequations}
\label{changerepresentionS}
\begin{align}
{\cal S} &= {\cal W}^{\dagger} s {\cal W} \, , \label{changerepresentionS2} \\
S^{00}    &= s^{00} + \frac12 \left[\Tr\left(s^{11}\right) -  \Tr\left(S^{11}\right)\right]  \, , \label{changerepresentionS1} \end{align}
\end{subequations}
and can be obtained through specific sets of expectation values
\begin{subequations}
\label{expect2}
\begin{align}
 s^{00} &= \elma{0}{S}{0} \, , \\
 s^{11}_{pq} &= \elma{0}{\left[ c_{p} , S\right] c_{q}^\dagger}{0} \, , \\
 s^{20}_{pq} &= \elma{0}{ c_{q} \left[ c_{p} ,S\right]}{0} \, , \\
 s^{02}_{pq} &= \elma{0}{[ S , c_{p}^\dagger ] \, c_{q}^\dagger }{0} \, ,
\end{align}
\end{subequations}
and
\begin{subequations}
\label{expect1}
\begin{align}
 S^{00} &= \elma{\Phi}{S}{\Phi} \, , \\
 S^{11}_{k_1 k_2} &= \elma{\Phi}{\left[ \beta_{k_1} , S\right] \beta_{k_2}^\dagger}{\Phi} \, , \\
 S^{20}_{k_1 k_2} &= \elma{\Phi}{ \beta_{k_2} \left[ \beta_{k_1} ,S\right]}{\Phi} \, , \\
 S^{02}_{k_1 k_2} &= \elma{\Phi}{[ S , \beta_{k_1}^\dagger ] \, \beta_{k_2}^\dagger }{\Phi} \, .
\end{align}
\end{subequations}

\subsection{Bogoliubov transformation between both vacua}

As already alluded to, there exist two situations to be distinguished. The first one concerns the particular case where $\ket{\breve{\Phi}}$ is {\it explicitly} obtained from $\ket{\Phi}$ via a known unitary transformation of the form~\ref{defphibreve}. In this situation, the Bogoliubov transformation $\breve{{\cal W}}$ is deduced from ${\cal W}$ and the {\it known} components of $S$. This is for instance the case when $\ket{\breve{\Phi}}$ is obtained via a one-body symmetry transformation, e.g. via a global gauge transformation of $\ket{\Phi}$ within the frame of particle-number restoration calculations or via a three-dimensional space rotation of $\ket{\Phi}$ within the frame of angular-momentum restoration calculations. The second, more general, situation relates to the case where $\ket{\breve{\Phi}}$ is not originally defined through such a transformation of $\ket{\Phi}$, i.e. it is obtained independently of $\ket{\Phi}$. It is what happens when $\ket{\Phi}$ and $\ket{\breve{\Phi}}$ are obtained, e.g., within the frame of a generator coordinate method calculation along an arbitrary collective coordinate. In this situation, ${\cal W}$ and $\breve{{\cal W}}$ are the given of the problem and $S$ has to be reconstructed from them a posteriori.

In either case, one eventually needs to represent the unitary Bogoliubov transformation linking the quasi-particle operators associated with $\ket{\Phi}$ and $\ket{\breve{\Phi}}$ via
\begin{equation}
\label{bogotransfolink}
\begin{pmatrix}
\beta \\
\beta^{\dagger}
\end{pmatrix}
 = {\cal W}^\dagger \breve{{\cal W}} 
\begin{pmatrix}
\breve{\beta} \\
\breve{\beta}^{\dagger}
\end{pmatrix}  
\equiv {\cal X}^{ \dagger} 
\begin{pmatrix}
\breve{\beta} \\
\breve{\beta}^{\dagger}
\end{pmatrix}  \, ,
\end{equation}
where
\begin{align}
{\cal X}
&\equiv \begin{pmatrix}
A & B^{\ast} \\
B & A^{\ast}
\end{pmatrix}  \nonumber \\
&= 
\begin{pmatrix}
\breve{U}^\dagger U + \breve{V}^\dagger V & \breve{U}^\dagger V^* + \breve{V}^\dagger U^*  \\
\breve{U}^T V + \breve{V}^T U & \breve{U}^T U^* + \breve{V}^T V^*
\end{pmatrix} \, .  \label{bogotransfolink2}
\end{align}
Thanks to the unitarity of ${\cal X}$, identities similar to those appearing in Eq.~\ref{unitarity} hold for matrices $A$ and $B$.

\subsection{Manifold of transformed states}

As expressed in Eq.~\ref{eq:genov2}, the computation of the norm kernel involves an integration along the auxiliary manifold $\mathcal{M}[| \Phi \rangle,S]$ introduced in Sec.~\ref{master}. Consequently, one needs to fully characterize the vacua constituting the manifold.

\subsubsection{Bogoliubov transformation between $\ket{\Phi(\theta)}$ and $\ket{\Phi}$}

The set of quasi-particle operators $\left\{  \beta^{\theta}_k ; \beta^{\theta +}_k\right\}$ associated with $| \Phi(\theta) \rangle$ are defined via
\begin{equation}
\label{tranfoBogotheta}
\begin{pmatrix}
\beta \\
\beta^{\dagger}
\end{pmatrix}
= {\cal X}(\theta)^{ \dagger} 
\begin{pmatrix}
\beta^{\theta} \\
\beta^{\theta \dagger}
\end{pmatrix}  \, ,
\end{equation}
where
\begin{equation}
{\cal X}(\theta) \equiv
\begin{pmatrix}
A(\theta) & B^{\ast}(\theta) \\
B(\theta) &  A^{\ast}(\theta)
\end{pmatrix} 
\, , \label{bogomani} 
\end{equation}
with the boundary conditions ${\cal X}(0)=1$ and ${\cal X}(1)={\cal X}$, i.e. 
\begin{equation}
A(0)=1   \, ; \, B(0)=0 \, , 
\end{equation}
and
\begin{align}
A(1)=A  \, ; \,  B(1)=B \, .
\end{align}

The Bogoliubov transformation ${\cal X}(\theta)$ is to be explicited in terms of the matrix ${\cal S}$ and the angle $\theta$. To do so, we transform the quasi-particle operators associated with $\ket{\Phi}$ through the unitary operator $e^{i\theta S}$
\begin{subequations}
\label{transforoperators}
\begin{align}
\beta^{\theta}_{k_1} & \equiv  e^{i\theta S} \beta_{k_1} e^{-i\theta S} \nonumber \\
&= \sum_{k_2} B^*_{k_1 k_2}(\theta) \beta^{\dagger}_{k_2} + A_{k_1 k_2}(\theta) \beta_{k_2} \, , \\
\beta^{\theta \dagger}_{k_1} & \equiv  e^{i\theta S} \beta^{\dagger}_{k_1} e^{-i\theta S} \nonumber \\
&= \sum_{k_2} A^*_{k_1 k_2}(\theta) \beta^{\dagger}_{k_2} + B_{k_1 k_2}(\theta) \beta_{k_2} \, .
\end{align}
\end{subequations}
To obtain an explicit representation of ${\cal X}(\theta)$, one takes the derivative\footnote{For any matrix $M(\theta)$, we use the standard notation $\frac{d}{d\theta}M(\theta)\equiv M^{\prime}(\theta)$ throughout the paper.} of Eq.~\ref{transforoperators} with respect to $\theta$ before setting $\theta=0$~\cite{hara79a}
\begin{subequations}
\label{derivativebogo}
\begin{align}
 i \left[ S , \beta_{k_1} \right]  &= \sum_{k_2} B^{\prime*}_{k_1 k_2}(0) \beta^{\dagger}_{k_2} + A^{\prime}_{k_1 k_2}(0) \beta_{k_2} \, , \\
 i \left[ S , \beta^{\dagger}_{k_1} \right]  &= \sum_{k_2} A^{\prime*}_{k_1 k_2}(0) \beta^{\dagger}_{k_2} + B^{\prime}_{k_1 k_2}(0) \beta_{k_2} \, .
\end{align}
\end{subequations}
Taking expectation values according to Eq.~\ref{expect1} leads to
\begin{subequations}
\begin{align}
 A^{\prime}_{k_1 k_2}(0) &= - i S^{11}_{k_1 k_2} \, , \\
 B^{\prime}_{k_1 k_2}(0) &= + i S^{02}_{k_1 k_2} \, ,
\end{align}
\end{subequations}
such that
\begin{equation}
{\cal X}^\prime(0) = -i{\cal S} \, .
\end{equation}
Next, one exploits that the unitary transformations underlying the manifold are abelian, as $S$ commutes with itself, to write
\begin{equation}
 {\cal X}(\theta_1 + \theta_2) = {\cal X}(\theta_1) {\cal X}(\theta_2) = {\cal X}(\theta_2) {\cal X}(\theta_1) \, .
\end{equation}
Taking the derivative with respect to $\theta_2$ before setting $\theta_1=\theta$ and $\theta_2=0$ leads to the first-order differential equation
\begin{equation}
 \label{explicitformtransfo}
 {\cal X}^\prime(\theta) = -i{\cal X}(\theta) \, {\cal S} = -i{\cal S} \, {\cal X}(\theta) \, ,
\end{equation}
whose solution provides an exponential representation of ${\cal X}(\theta)$ in terms of ${\cal S}$ and $\theta$
\begin{equation}
 {\cal X}(\theta) =  e^{-i\theta {\cal S}} \label{exprepresentationWtheta}
 \, .
\end{equation}

Given the hermitian character of ${\cal S}$, ${\cal X}(\theta)$ is indeed unitary and qualifies as a Bogoliubov transformation between the initial and the transformed quasi-particle operators. Furthermore, it is straightforward to see that
\begin{equation}
\beta^{\theta}_{k_1} \ket{\Phi(\theta)} = e^{i\theta S} \beta_{k_1} \ket{\Phi} = 0
\end{equation}
for all $k_1$, such that $\ket{\Phi(\theta)}$ is a vacuum for the set of transformed quasi-particle operators $\left\{  \beta^{\theta}_k ; \beta^{\theta +}_k\right\}$. For $\theta =1$, one further accesses the exponential representation of the Bogoliubov transformation ${\cal X}$ linking $\ket{\Phi}$ to $\ket{\breve{\Phi}}$ (Eq.~\ref{bogotransfolink2}).

As $S^{11}$ is hermitian, one has 
\begin{equation}
\Tr{\cal S} = \Tr
\begin{pmatrix}
 S^{11} &  S^{20} \\
 -S^{02} & - S^{11*}
\end{pmatrix} 
= 0 \, , 
\end{equation}
such that, by virtue of Jacobi's formula\footnote{The proof is trivial if the exponentiated matrix is diagonalizable as  is presently the case.}, the representation provided in Eq.~\ref{exprepresentationWtheta} is characterized by
\begin{equation}
\det{\cal X}(\theta)=  e^{-i\theta \Tr{\cal S}} = +1 \, , \label{detW}
\end{equation}
which is consistent with the fact that ${\cal X}(\theta)$ is a proper Bogoliubov transformation.

\subsubsection{Practical extraction of ${\cal X}(\theta)$: simple case}
\label{extract1}

Whenever the given of the problem are the Bogoliubov transformation ${\cal W}$ and the operator $S$, ${\cal X}(\theta)$ can be computed straightforwardly on the basis of Eq.~\ref{exprepresentationWtheta}.  To proceed, one diagonalizes the hermitian matrix ${\cal S}$ according to 
\begin{equation}
{\cal P}^{\dagger}
\begin{pmatrix}
 S^{11} &  S^{20} \\
 -S^{02} & - S^{11*}
\end{pmatrix} 
{\cal P}
\equiv
{\cal S}_{\text{D}}  \, , \label{diago1}
\end{equation}
where the eigenvalues on the diagonal of the $2N \times 2N$ matrix ${\cal S}_{\text{D}}$ are real numbers
\begin{eqnarray}
{\rm Sp} \, {\cal S} &=& \{s_i \in \mathbbm{R},  i =1, \ldots N ;\\
&&\,\, s_j \equiv -s_{j-N},  j =N+1, \ldots 2N\} \, . \nonumber
\end{eqnarray}
According to Eq.~\ref{exprepresentationWtheta}, ${\cal X}(\theta)$ is simultaneously diagonal with eigenvalues $x_i(\theta)$ on the diagonal of ${\cal X}_{\text{D}}(\theta)$ of the form
\begin{equation}
{\rm Sp} \, {\cal X}(\theta) = \{x_i(\theta) = e^{-i \theta s_i},  i =1, \ldots 2N\} \, , \nonumber
\end{equation}
such that
\begin{equation}
{\cal X}(\theta) =
\begin{pmatrix}
A(\theta) & B(\theta)^{\ast}  \\
B(\theta) &  A(\theta)^{\ast}
\end{pmatrix} 
=
 {\cal P}
{\cal X}_{\text{D}}(\theta){\cal P}^{\dagger}
 \, . \label{reconstruct1}
\end{equation}
For $\theta =1$, Eq.~\ref{reconstruct1} provides the Bogoliubov transformation ${\cal X}$ linking $\ket{\Phi}$ and $\ket{\breve{\Phi}}$ introduced in Eq.~\ref{bogotransfolink2}, further leading to the Bogoliubov transformation defining $\ket{\breve{\Phi}}$ itself via
\begin{equation}
\breve{{\cal W}} 
\equiv {\cal W} {\cal X}^{ \dagger} 
  \, .
\end{equation}

\subsubsection{Practical extraction of ${\cal X}(\theta)$: general case}
\label{extract2}

Whenever the given of the problem are the Bogoliubov transformations ${\cal W}$ and $\breve{{\cal W}}$, the situation is more involved. One first needs to extract the elements defining the matrix ${\cal S}$ before obtaining ${\cal X}(\theta)$ as above. 

Since $\ket{\Phi(1)} = \ket{\breve{\Phi}}$, one first exploits the boundary condition ${\cal X}(1)={\cal X}$ writing as
\begin{equation}
\exp\left\{-i
\begin{pmatrix}
 S^{11} &  S^{20} \\
 -S^{02} & - S^{11*}
\end{pmatrix} 
\right\} = \begin{pmatrix}
A & B^{\ast} \\
B & A^{\ast}
\end{pmatrix} 
 \, , \label{matching}
\end{equation}
to determine matrices $S^{11}$, $S^{20}$ and $S^{02}$. As ${\cal X}$ is unitary, it can be diagonalized according to
\begin{equation}
{\cal P}^{\dagger}\begin{pmatrix}
A & B^{\ast} \\
B & A^{\ast}
\end{pmatrix} 
{\cal P}
\equiv
{\cal X}_{\text{D}}   \, , \label{diago2}
\end{equation}
where the $2N$ eigenvalues are unitary complex numbers satisfying
\begin{eqnarray}
{\rm Sp} \, {\cal X} &=& \{x_i \, / \, |x_i|=1,  i =1, \ldots N ;\\
&&\,\, x_j \equiv x^{\ast}_{j-N},  j =N+1, \ldots 2N\} \, . \nonumber
\end{eqnarray}
Representing these eigenvalues as
\begin{equation}
x_i \equiv e^{-i s_i} 
\end{equation}
with $s_i \in ]-\pi,\pi]$ for $i =1, \ldots 2N$, ${\cal S}$ is simultaneously diagonal with eigenvalues obtained via the principal logarithm\footnote{The reason why the principal logarithm is the only viable option is discussed in App.~\ref{unicityS}. Despite the a priori multivalued character of the logarithm of a complex matrix, matrix ${\cal S}$ is thus uniquely defined in the present context.} of ${\cal X}_{\text{D}}$
\begin{equation}
{\rm Sp} \, {\cal S} \equiv \{s_i = i\log x_i \in  ]-\pi,\pi],  i =1, \ldots 2N\} \, . \nonumber
\end{equation}
This allows one to extract matrices $S^{11}$, $S^{20}$ and $S^{02}$ according to
\begin{equation}
\begin{pmatrix}
 S^{11} &  S^{20} \\
 -S^{02} & - S^{11*}
\end{pmatrix} 
=
 {\cal P}
{\cal S}_{\text{D}}{\cal P}^{\dagger}
 \, , \label{reconstruct2}
\end{equation}
and, following the development provided in Sec.~\ref{extract1}, to obtain ${\cal X}(\theta)$ through 
\begin{equation}
{\cal X}(\theta) =
\begin{pmatrix}
A(\theta) & B(\theta)^{\ast} \nonumber \\
B(\theta) &  A(\theta)^{\ast}
\end{pmatrix} 
=
 {\cal P}
{\cal X}_{\text{D}}(\theta){\cal P}^{\dagger}
 \, , \label{reconstruct3}
\end{equation}
where ${\cal X}_{\text{D}}(\theta)$ is the diagonal matrix with entries 
\begin{equation}
{\rm Sp} \, {\cal X}(\theta) = \{x_i(\theta) = e^{-i \theta s_i},  i =1, \ldots 2N\} \, . \nonumber
\end{equation}

\subsubsection{Off-diagonal contractions}
\label{subsectelemcont}

With matrices $A(\theta)$ and $B(\theta)$ at hand, one is in position to compute elementary off-diagonal contractions between $\ket{\Phi(\theta)}$ and $\ket{\Phi}$ that are eventually needed to calculate the off-diagonal kernel of $S$ in Eq.~\ref{eq:genovHFBbis}. These are given by
\begin{eqnarray}
{\cal R}[\langle \Phi |,| \Phi(\theta) \rangle] &\equiv& 
\left(
\begin{array} {cc}
R^{+-}[\langle \Phi |,| \Phi(\theta) \rangle] & R^{--}[\langle \Phi |,| \Phi(\theta) \rangle] \\
R^{++}[\langle \Phi |,| \Phi(\theta) \rangle] &  R^{-+}[\langle \Phi |,| \Phi(\theta) \rangle]
\end{array}
\right) \nonumber  \\
&\equiv& 
\left(
\begin{array} {cc}
\frac{\langle \Phi | \beta^{\dagger}\beta^{\phantom{\dagger}} | \Phi(\theta) \rangle}{\langle \Phi | \Phi(\theta) \rangle} & \frac{\langle \Phi | \beta^{\phantom{\dagger}}\beta^{\phantom{\dagger}} | \Phi(\theta) \rangle}{\langle \Phi | \Phi(\theta) \rangle} \\
\frac{\langle \Phi | \beta^{\dagger}\beta^{\dagger} | \Phi(\theta) \rangle}{\langle \Phi | \Phi(\theta) \rangle} &  \frac{\langle \Phi | \beta^{\phantom{\dagger}}\beta^{\dagger} | \Phi(\theta) \rangle}{\langle \Phi | \Phi(\theta) \rangle}
\end{array}
\right)  \nonumber \\
&=& 
\left(
\begin{array} {cc}
0 & -B^{\dagger}(\theta)[A^{T}(\theta)]^{-1} \\
0 &  1
\end{array}
\right)
\, , \label{offdiaggeneralizeddensitymatrix2}
\end{eqnarray}
whose components are trivial except for $R^{--}[\langle \Phi |,| \Phi(\theta) \rangle]$. This results from the fact that the contractions are defined in the quasi-particle basis associated with the bra, i.e. $\langle \Phi |$ here. The explicit form of $R^{--}[\langle \Phi |,| \Phi(\theta) \rangle]$ is obtained by transforming the contractions computed in the single-particle basis (see, e.g., App. E of Ref.~\cite{ring80a}) into the quasi-particle basis. The skew-symmetry of $R^{--}[\langle \Phi |,| \Phi(\theta) \rangle]$ reads as
\begin{subequations}
\label{upperright}
\begin{align}
R^{--}[\langle \Phi |,| \Phi(\theta) \rangle]&= - R^{-- T}[\langle \Phi |,| \Phi(\theta) \rangle]  \\
&= A^{-1}(\theta)B^{*}(\theta) \, . 
\end{align}
\end{subequations}
Given that $A(0)=1$ and $B(0)=0$, one further notices that $R^{--}[\langle \Phi |,| \Phi(0) \rangle]=R^{--}[\langle \Phi |,| \Phi \rangle]=0$.

\subsection{Norm kernel}

\subsubsection{Generic expression}

Inserting Eq.~\ref{eq:1bodynobogo} into Eq.~\ref{eq:genov} at $\tau=0$ the uncorrelated norm kernel along the auxiliary manifold is expressed, on the basis of the off-diagonal Wick's theorem~\cite{balian69a}, as
\begin{align}
\frac{\langle \Phi | \Phi(\theta) \rangle}{\langle \Phi | \Phi \rangle} &=  e^{i \theta S^{00}} e^{ \frac{i}{2} \sum_{k_1 k_2} S^{02}_{k_1 k_2} \int_0^\theta d\phi \, R^{--}_{k_2 k_1}[\langle \Phi |,| \Phi(\phi) \rangle]} \nonumber \\  
                            &=  e^{i \theta S^{00}} e^{ \frac{i}{2} \int_0^\theta d\phi \Tr \left( S^{02} R^{--}[\langle \Phi |,| \Phi(\phi) \rangle] \right)} \, , \label{eq:genovHFBNOgeneric}
\end{align}
which, thanks to $R^{--}[\langle \Phi |,| \Phi(0) \rangle]=0$, leads in particular to 
\begin{equation}
S^{00} = -i \left[ \frac{d}{d\theta}  \frac{\langle \Phi | \Phi(\theta) \rangle}{\langle \Phi | \Phi \rangle}\right]_{\theta = 0} \, . \label{zeroderivative}
\end{equation}

Taking $\theta=1$ in Eq.~\ref{eq:genovHFBNOgeneric} eventually reexpresses Eq.~\ref{eq:genovHFB} under the workable form
\begin{equation}
\label{eq:genovHFBNO}
\begin{split}
\frac{\langle \Phi | \breve{\Phi} \rangle}{\langle \Phi | \Phi \rangle} &=  e^{i S^{00}} e^{ \frac{i}{2} \int_0^1 d\theta \Tr \left( S^{02} R^{--}[\langle \Phi |,| \Phi(\theta) \rangle] \right)} \, ,
\end{split}
\end{equation}
with $R^{--}[\langle \Phi |,| \Phi(\theta) \rangle]$ expressed in terms of matrices $A(\theta)$ and $B(\theta)$ (Eq.~\ref{offdiaggeneralizeddensitymatrix2}), themselves related to components of ${\cal S}$. Two comments are in order

\begin{itemize}
\item As visible from Eq.~\ref{eq:genovHFBNO}, $S^{02}$ is key to capturing the norm kernel, highlighting the necessity to build the unitary transformation relating $| \breve{\Phi} \rangle$ to $| \Phi \rangle$. Conversely, the norm kernel cannot be obtained from the non-unitary Thouless transformation between both vacua given that the operator driving this transformation contains an operator proportional to two quasi-particle creation operators (i.e. similar to $S^{20}$) but not its hermitian congugate (i.e. similar to $S^{02}$).
\item An important assumption to derive Eq.~\ref{eq:genovHFBNO} is that $| \Phi \rangle$ and $| \Phi(\theta) \rangle$ are not orthogonal over the interval $\theta \in [0,1]$. It is in fact not necessary to assume this property for $\theta=1$ as the orthogonality of $| \breve{\Phi} \rangle$ and $| \Phi \rangle$ is well captured at the price of seeing Eq.~\ref{eq:genovHFBNO} as the {\it limit} of Eq.~\ref{eq:genovHFBNOgeneric} for $\theta \rightarrow 1$, whenever the real part of the argument of the exponential goes to $-\infty$. This point is analytically scrutinized in App.~\ref{sec:divergence} and numerically illustrated in Sec.~\ref{resultsgaugerotation} in the case of global gauge transformations.
\end{itemize}

\subsubsection{Phase convention}

As visible from Eq.~\ref{exprepresentationWtheta}, the Bogoliubov transformation ${\cal X}(\theta)$ built from ${\cal W}$ and $\breve{{\cal W}}$ is insensitive to the pure number $S^{00}$ entering $S$ and Eq.~\ref{eq:genovHFBNO}. Conversely, $S^{00}$ cannot be determined from ${\cal X}$, which relates to the fact that a phase convention must be chosen to fix the associated freedom. This has already been translated into the condition manifested by Eq.~\ref{phasecondition1}.

In the present section, Eq.~\ref{phasecondition1} is worked out in the case where the common reference state is chosen to be the particle vacuum, i.e. when setting $\ket{\bar{\Phi}}=\ket{0}$. The generalization to any appropriate $| \bar{\Phi} \rangle$ is straightforward. We introduce the Bogoliubov transformation linking the operators associated with $\ket{\Phi(\theta)}$ and $\ket{0}$
\begin{equation}
\begin{pmatrix}
c \\
c^{\dagger}
\end{pmatrix}
\equiv  {\cal Y}(\theta)^{ \dagger} 
\begin{pmatrix}
\beta^{\theta} \\
\beta^{\theta \dagger}
\end{pmatrix} 
 \, ,
\end{equation}
with\footnote{Exploiting Eq.~\ref{exprepresentationWtheta}, one obtains the exponential representation of ${\cal Y}(\theta)$ under the form
\begin{equation}
{\cal Y}(\theta) = e^{-i\theta {\cal S}} {\cal W}^{\dagger} = {\cal W}^{\dagger} e^{-i\theta s}  \, ,
\end{equation}
where $s={\cal W} {\cal S} {\cal W}^{\dagger}$ was used.}
\begin{align}
{\cal Y}(\theta) 
 &= 
 {\cal X}(\theta) {\cal W}^{ \dagger} \nonumber \\
 &= 
 \begin{pmatrix}
  A(\theta)U^\dagger  + B^*(\theta)V^T & B^*(\theta)U^T  + A(\theta)V^\dagger \\
  B(\theta)U^\dagger  + A^*(\theta)V^T & A^*(\theta)U^T  + B(\theta)V^\dagger \\
 \end{pmatrix}  
 \nonumber \\
 &\equiv
 \begin{pmatrix}
  C(\theta) & D^*(\theta)\\
  D(\theta) & C^*(\theta)
 \end{pmatrix}  \, ,
\end{align}
and the boundary conditions ${\cal Y}(0) ={\cal W}^{\dagger}$ and ${\cal Y}(1) =\breve{{\cal W}}^{\dagger}$. 

The elementary off-diagonal contractions between both vacua are given by
\begin{eqnarray}
{\cal R}[\langle 0 |,| \Phi(\theta) \rangle] &\equiv& 
\left(
\begin{array} {cc}
R^{+-}[\langle 0 |,| \Phi(\theta) \rangle] & R^{--}[\langle 0 |,| \Phi(\theta) \rangle] \\
R^{++}[\langle 0 |,| \Phi(\theta) \rangle] & R^{-+}[\langle 0 |,| \Phi(\theta) \rangle]
\end{array}
\right)  \nonumber \\
&\equiv& 
\left(
\begin{array} {cc}
\frac{\langle 0 | c^{\dagger}c^{\phantom{\dagger}} | \Phi(\theta) \rangle}{\langle 0 | \Phi(\theta) \rangle} & \frac{\langle 0 | c^{\phantom{\dagger}}c^{\phantom{\dagger}} | \Phi(\theta) \rangle}{\langle 0 | \Phi(\theta) \rangle} \\
\frac{\langle 0 | c^{\dagger}c^{\dagger} | \Phi(\theta) \rangle}{\langle 0 | \Phi(\theta) \rangle} &  \frac{\langle 0 | c^{\phantom{\dagger}}c^{\dagger} | \Phi(\theta) \rangle}{\langle 0 | \Phi(\theta) \rangle}
\end{array}
\right)  \nonumber \\
&=& 
\left(
\begin{array} {cc}
0 & -D^{\dagger}(\theta)[C^{T}(\theta)]^{-1} \\
0 &  1
\end{array}
\right)
 \nonumber \\
&=& 
\left(
\begin{array} {cc}
0 & C^{-1}(\theta)D^{*}(\theta) \\
0 &  1
\end{array}
\right)
\, , \label{offdiaggeneralizeddensitymatrix2}
\end{eqnarray}
and are related to those introduced earlier via
\begin{eqnarray}
R^{--}[\langle 0 |,| \Phi(\theta) \rangle] 
&=& - \left[ V^* - U R^{--}[\langle \Phi |,| \Phi(\theta) \rangle] \right] \nonumber \\
&& \times \left[ U^* - V R^{--}[\langle \Phi |,| \Phi(\theta) \rangle] \right]^{-1} \, . \nonumber \\
\end{eqnarray}

With these ingredients at hand, Eq. ~\ref{phasecondition1} rewrites as
\begin{align}
s^{00} &= - \Re e \frac12 \int_0^1 d\theta \, \Tr\left(s^{02} R^{--}[\langle 0 |,| \Phi(\theta) \rangle]\right)  \, ,
\end{align}
which, thanks to Eq.~\ref{changerepresentionS1}, can be trivially transformed as an equation for $S^{00}$. Eventually,  the final expression of the uncorrelated overlap is
\begin{eqnarray}
\frac{\langle \Phi | \breve{\Phi} \rangle}{\langle \Phi | \Phi \rangle} &=&  e^{\frac{i}{2} \left[\Tr\left(s^{11}\right)-  \Re e \int_0^1 d\theta \Tr \left(s^{02} R^{--}[\langle 0 |,| \Phi(\theta) \rangle]\right) \right]} \nonumber \\
&\times& e^{-\frac{i}{2} \left[\Tr\left(S^{11}\right)-  \int_0^1 d\theta \Tr \left(S^{02} R^{--}[\langle \Phi |,| \Phi(\theta) \rangle]\right)\right]} \, , \label{overlapreformulated}
\end{eqnarray}
where all ingredients are known on the sole basis of ${\cal W}$ and $\breve{{\cal W}}$.

\subsubsection{Algorithm}
\label{algorithm}

In summary, the steps to compute the norm kernel between $\ket{\breve{\Phi}}$ and $\ket{\Phi}$, with ${\cal W}$ and $\breve{{\cal W}}$ as the sole inputs, are
\begin{enumerate}
\item Compute the Bogoliubov matrix ${\cal X}=\breve{{\cal W}}^{\dagger}{\cal W}$.
\item Diagonalize the unitary matrix ${\cal X}$ to extract 
\begin{enumerate}
\item ${\cal S}=i \log {\cal X}$,
\item $s={\cal W}{\cal S}{\cal W}^{\dagger}$,
\item ${\cal X}(\theta)=\exp( -i\theta{\cal S})$,
\item ${\cal Y}(\theta)={\cal X}(\theta){\cal W}^{\dagger}$.
\end{enumerate}
\item Compute 
\begin{enumerate}
\item $R^{--}[\langle \Phi |,| \Phi(\theta) \rangle]=A^{-1}(\theta)B^{*}(\theta)$,
\item $R^{--}[\langle 0 |,| \Phi(\theta) \rangle]=C^{-1}(\theta)D^{*}(\theta)$,
\end{enumerate}
from submatrices of ${\cal X}(\theta)$ and ${\cal Y}(\theta)$, respectively.
\item Compute the norm kernel $\langle \Phi | \breve{\Phi} \rangle/\langle \Phi | \Phi \rangle$ via Eq.~\ref{overlapreformulated}.
\end{enumerate}
Thus, the procedure involves matrix multiplications, the diagonalization of a unitary matrix, computing the inverse of two (potentially singular) matrices and performing an integral.

\subsubsection{Connection to the Onishi formula}
\label{rewritingoverlap}

Equation~\ref{overlapreformulated} is a workable expression of the uncorrelated overlap kernel. It involves traces and numerical integrations of products of known matrices running over the manifold of states $\mathcal{M}[| \Phi \rangle,S]$. In that sense, Eq.~\ref{overlapreformulated} explicitly follows a unitary path from $| \Phi \rangle$ to $| \breve{\Phi} \rangle$ by integrating over the manifold. The well-known Onishi formula~\cite{onishi66}, on the other hand, solely provides the norm of the overlap kernel by expressing it in terms of the matrix $A$ involved in the Bogoliubov transformation ${\cal X}$ linking $| \Phi \rangle$ and $| \breve{\Phi} \rangle$ directly, i.e. jumping over the continuous path going from the bra to the ket, it looses the phase of the overlap. 

To recover the Onishi formula, one starts by left-multiplying the upper-left corner of Eq.~\ref{explicitformtransfo} to obtain~\cite{hara79a}
\begin{align}
i \Tr \left(S^{02} R^{--}[\langle \Phi |,| \Phi(\theta) \rangle]\right) - i \Tr\left(S^{11}\right)  & = \Tr\left(A^{-1}(\theta)A^{\prime}(\theta)\right) \nonumber \\
&= \frac{d}{d\theta}  \Tr \left(\ln A(\theta)\right) \, . \label{haraidentities}
\end{align}
Considering the norm of Eq.~\ref{overlapreformulated} and inserting Eq.~\ref{haraidentities} before proceeding to the integration under the conditions $A(0)=1$ and $A(1)=A$ leads to the Onishi formula
\begin{align}
\left|\frac{\langle \Phi | \breve{\Phi} \rangle}{\langle \Phi | \Phi \rangle}\right| & =  \left|e^{\frac{1}{2} \Tr \left(\ln A \right)}\right|  = \sqrt{\left|\det A\right|} \, . \label{onishi}
\end{align}

\subsection{Family of equivalent auxiliary manifolds}
\label{family}

Given the Bogoliubov transformation ${\cal W}$, let us define a new set of quasiparticle operators via\footnote{The procedure described in this section can be equally applied to $\breve{{\cal W}}$.}
\begin{subequations}
\label{furthertransfo1}
\begin{align}
\begin{pmatrix}
\tilde{\beta} \\
\tilde{\beta}^{\dagger}
\end{pmatrix}
&\equiv \tilde{{\cal W}}^{\dagger} 
\begin{pmatrix}
c \\
c^{\dagger}
\end{pmatrix}  \\
&= {\cal K}^{\dagger} 
\begin{pmatrix}
\beta \\
\beta^{\dagger}
\end{pmatrix}  \, , \label{furthertransfo1B}
\end{align}
\end{subequations}
with ${\cal K}$ a trivial Bogoliubov transformation, i.e. 
\begin{subequations}
\label{furthertransfo2}
\begin{align}
\tilde{{\cal W}} &\equiv {\cal W}{\cal K} \\
&= 
\begin{pmatrix}
U & V^{\ast} \\
V &  U^{\ast}
\end{pmatrix}  
\begin{pmatrix}
K & 0 \\
0 &  K^{\ast}
\end{pmatrix}  
 \\
&=  \begin{pmatrix}
UK & V^{\ast}K^{\ast} \\
VK &  U^{\ast}K^{\ast} 
\end{pmatrix}  
 \\
&\equiv  \begin{pmatrix}
\tilde{U} & \tilde{V}^{\ast} \\
\tilde{V} &  \tilde{U}^{\ast} 
\end{pmatrix}  
 \,  , \label{newqp}
\end{align}
\end{subequations}
where $K$ is a $N\times N$ unitary matrix. Defined in this way, the new set of quasiparticle creation (annihilation) operators result from a transformation of the original creation (annihilation) quasiparticle operators among themselves\footnote{This constitutes a trivial Bogoliubov transformation of the same type as the third transformation in the Bloch-Messiah-Zumino decomposition of a non-trivial Bogoliubov transformation~\cite{ring80a}. }, i.e. 
\begin{subequations}
\begin{align}
 \tilde{\beta}_{k_1} &= \sum_{k_2} \tilde{U}^*_{k_2 k_1} \, c_{k_2} + \tilde{V}^*_{k_2 k_1} \, c_{k_2}^\dagger   \\
 					 &= \sum_{k_3} K^*_{k_3 k_1} \, \beta_{k_3} \, , \\
 \tilde{\beta}_{k_1}^\dagger &= \sum_{k_2} \tilde{U}_{k_2 k_1} \, c_{k_2}^\dagger + \tilde{V}_{k_2 k_1} \, c_{k_2} \\ 
 							 &= \sum_{k_3} K_{k_3 k_1} \, \beta_{k_3}^\dagger  \, .  
\end{align}
\end{subequations}
The Bogoliubov state $| \Phi \rangle$ is also a vacuum for the new set of operators  $\{\tilde{\beta}_k;\tilde{\beta}^{\dagger}_k\}$ given that
\begin{align}
\tilde{\beta}_{k_1} \ket{\Phi} &=  \sum_{k_3} K^*_{k_3 k_1} \, \beta_{k_3} \ket{\Phi} = 0
\end{align}
for all $k_1$. 

If one chooses to explicitly represent the vacuum state in terms of the original set $\{\beta_k;\beta^{\dagger}_k\}$ via, e.g.,
\begin{equation}
| \Phi \rangle \equiv \prod_{k_1} \beta_{k_1} | 0 \rangle \, ,
\end{equation}
one can easily prove that the state associated with the new set solely differs from the original one by a phase, i.e.
\begin{align}
| \tilde{\Phi} \rangle &\equiv \prod_{k_1} \tilde{\beta}_{k_1} | 0 \rangle \nonumber \\
&= \det K \, | \Phi \rangle \, ,
\end{align}
with $\det K \equiv e^{i\alpha}$ as $K$ is unitary. Consequently, the overlap of interest changes accordingly
\begin{equation}
\frac{\langle \tilde{\Phi}  | \breve{\Phi} \rangle}{\langle \tilde{\Phi} | \tilde{\Phi} \rangle} 
= \left(\det K\right)^{\ast} \, \frac{\langle \Phi | \breve{\Phi} \rangle}{\langle \Phi | \Phi \rangle} \, .
\end{equation}
Contrarily, the present approach to the off-diagonal norm kernel does {\it not} rely on an explicit representation of the Bogoliubov states involved. Rather, only the unitary operator {\it linking} the two states is constructed explicitly under the requirement that both states entertain the same phase (whatever its value) with a given state of reference. Such a procedure ensures that the arbitrary transformation ${\cal K}$ of the form given by Eq.~\ref{furthertransfo1B} leaves, by construction, the norm kernel invariant. However, this remarkable result is obtained while modifying {\it non trivially} the Bogoliubov transformation ${\cal X}$, the matrix ${\cal S}$, the operator $S$ and thus the manifold $\mathcal{M}[| \Phi \rangle,S]$ connecting $| \Phi \rangle$ to $| \breve{\Phi} \rangle$. Eventually, this means that $\langle \Phi | \breve{\Phi} \rangle$ is left invariant while the intermediate values $\langle \Phi | \Phi(\theta) \rangle$ along the manifold are changed essentially at will. The practical benefit of proceeding to such harmless Bogoliubov transformations will be illustrated on the basis of the numerical applications discussed in Sec.~\ref{results}.

\section{Numerical results}
\label{results}

We now illustrate numerically the capacity of the exponential formula derived in the present work to efficiently capture the norm overlap between arbitrary Bogoliubov states. To do so, we employ several toy models of increasing complexity. In the first two models, reference values are provided by analytical formulae that can be straightforwardly derived without phase ambiguity. The generality of the last two calculations, however, makes only possible to benchmark the present approach against the Pfaffian method~\cite{Robledo:2009yd,Robledo:2011ce,Avez:2011wr}.

\subsection{Gauge transformation in a toy BCS model}
\label{resultsgaugerotation}

\begin{figure}[b]
\begin{center}
\includegraphics[width=3.5cm]{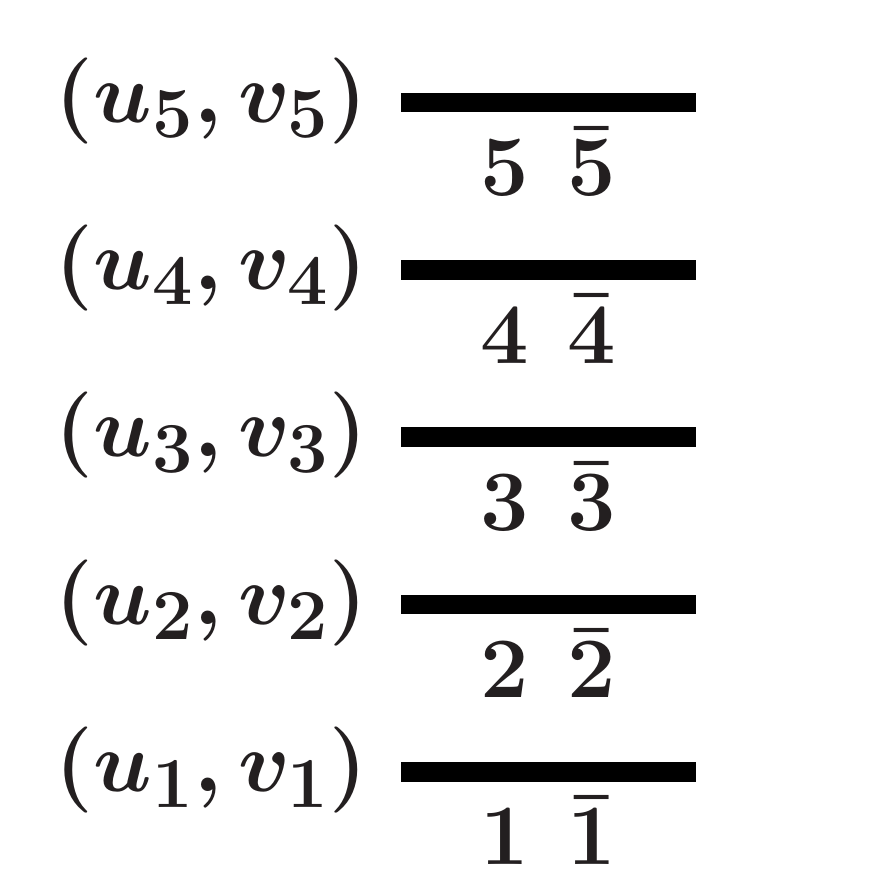}
\end{center}
\caption{Schematic representation of the five doubly degenerate single-particle levels from which a BCS state and its gauge-rotated partner are built.}
\label{10levelsBCSmodel}
\end{figure}

Our first numerical application deals with the overlap between two BCS states built out of five doubly-degenerated levels (see Fig.~\ref{10levelsBCSmodel}) and differing by a global gauge rotation. This constitutes the simplest situation in which the unitary transformation linking $| \Phi \rangle$ and $| \breve{\Phi} \rangle$ is known a priori, i.e. $S=\varphi A$.

\begin{figure}
\begin{center}
\includegraphics[width=6.5cm]{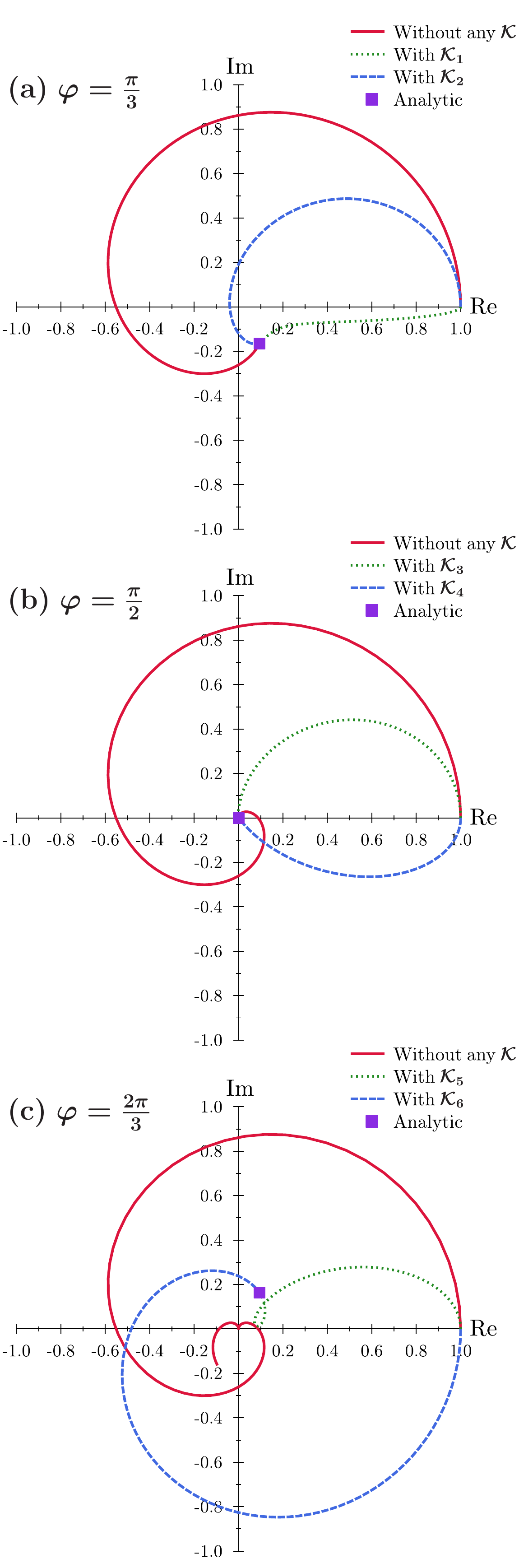}
\end{center}
\caption{(Color online) Norm overlap between the BCS state $| \Phi \rangle$ and its gauge-rotated partner $| \breve{\Phi} \rangle$ represented in the complex plane. The purple square denotes the reference result obtained from Eq.~\ref{referenceoverlapBCSgauge}. The color lines correspond to the increment integration along the auxiliary manifolds linking $| \Phi \rangle$ to $| \breve{\Phi} \rangle$ obtained without (full red line) or with (dashed blue and dotted green lines) an additional trivial Bogoliubov transformation ${\cal K}_i$ (see Sec.~\ref{family}). Upper panel: gauge angle $\varphi=\pi/3$. Middle panel: gauge angle $\varphi=\pi/2$. Lower panel: gauge angle $\varphi=2\pi/3$ In each panel, two different trivial Bogoliubov transformations ${\cal K}_i$ are randomly generated and used.}
\label{results10levelsBCSmodel}
\end{figure}

While we refer to App.~\ref{gaugerotation} for extensive analytical details regarding the norm overlap between Bogoliubov states differing by a gauge rotation, we specify here that the BCS transformation associated with $| \Phi \rangle$  is characterized by the set of real $2\times 2$ blocks of the form\footnote{The notation $(\bar{U},\bar{V})$ is presently used to specify that the Bogoliubov transformations at play are of BCS type but is unrelated to the generic notation $| \bar{\Phi} \rangle$ employed to denote the Bogoliubov state with respect to which the phase convention is set.}
\begin{subequations}
\label{BCStransforphi}
\begin{align}
\bar{U}(k,\bar{k}) &\equiv
\begin{pmatrix}
+u_k & 0 \\
0 &  +u_k
\end{pmatrix} \,\, , \\
\bar{V}(k,\bar{k}) &\equiv
\begin{pmatrix}
0 & +v_k \\
-v_k &  0
\end{pmatrix}  \,\, ,
\end{align}
\end{subequations}
with $\bar{k}$ denoting the conjugated partner of $k$ and with $u_k^2+v_k^2=1$. Correspondingly, the BCS transformation associated with $| \breve{\Phi} \rangle \equiv e^{i\varphi A} | \Phi \rangle$ is given by 
\begin{subequations}
\label{BCStransforphibreve}
\begin{align}
\breve{\bar{U}} &\equiv e^{+i\varphi} \bar{U} \,\, , \\
\breve{\bar{V}} &\equiv e^{-i\varphi} \bar{V} \,\, .
\end{align}
\end{subequations}
The two associated vacua can be explicitly represented by 
\begin{subequations}
\begin{align}
| \Phi \rangle &\equiv \prod_{k=1}^{5} (u_k + v_k c^{\dagger}_k c^{\dagger}_{\bar{k}}) | 0 \rangle \,\, , \\
| \breve{\Phi} \rangle &\equiv \prod_{k=1}^{5} (u_k + e^{2i\varphi} v_k c^{\dagger}_k  c^{\dagger}_{\bar{k}}) | 0 \rangle \,\, ,
\end{align}
\end{subequations}
which is consistent with the phase convention $\text{Arg} (\langle 0 | \Phi \rangle) = \text{Arg} (\langle 0 | \breve{\Phi} \rangle) = 0$. These two states are explicitly normalized and their complex overlap is easily shown to be
\begin{equation}
\frac{\langle \Phi | \breve{\Phi} \rangle}{\langle \Phi | \Phi \rangle}  = \prod_{k=1}^{5} (u^2_k + e^{2i\varphi} v^2_k ) \,\, , \label{referenceoverlapBCSgauge}
\end{equation}
which provides the formula of reference. 

The present toy model is fully defined once the single-particle occupations $v^2_k$ are specified for $k=1,\ldots 5$. In the numerical applications below, the occupations of the five doubly-degenerate levels are decreasingly chosen in the interval $ ]0,1[$ to qualitatively mimic a realistic fully paired system. The third level, in particular, is chosen to have $u^2_3=v^2_3=0.5$ in order to ensure that $\langle \Phi | \breve{\Phi} \rangle=0$ for $\varphi = \pi/2$.

Results are displayed in Fig.~\ref{results10levelsBCSmodel} for three representative values of the gauge angle. The  purple squares denote the reference values obtained from Eq.~\ref{referenceoverlapBCSgauge} while the lines characterize the increment integration along auxiliary manifolds linking $| \Phi \rangle$ to $| \breve{\Phi} \rangle$. The actual overlap of interest is thus the endpoint of these lines. The full red line is obtained by extracting the auxiliary manifold from the BCS transformations $\bar{{\cal W}}$ and $\breve{\bar{{\cal W}}}$ defined by Eqs.~\ref{BCStransforphi} and~\ref{BCStransforphibreve} without any further modification.  Consequently, the red path goes along the manifold of gauge rotated states (i.e. $S=\varphi A$) obtained for gauge angles $\phi \in [0,\varphi]$. Contrarily, the dashed blue and dotted green lines follow the auxiliary manifolds obtained by further multiplying $\breve{\bar{{\cal W}}}$ by two arbitrary\footnote{The trivial unitary Bogoliubov transformations ${\cal K}$ employed in the present calculations are randomly generated.} trivial Bogoliubov transformations ${\cal K}_i$ (see Sec.~\ref{family})\footnote{Even in the first case where the operator $S=\varphi A$ is  known a priori, we apply the  procedure outlined in Sec.~\ref{extract2} to extract $S$ based on the given of the BCS transformations associated with $| \Phi \rangle$ and $| \breve{\Phi} \rangle$. It allows us to check that the operator thus extracted is indeed nothing but $\varphi A$. In the other two cases, however, the additional transformation ${\cal K}$ makes the operator $S$ generating the unitary transformation between $| \Phi \rangle$ and $| \breve{\Phi} \rangle$ to be a priori unknown and eventually different from $\varphi A$.}.

Let us first focus on the upper panel corresponding to the gauge angle $\varphi=\pi/3$. We observe that the complex overlap is nicely captured by Eq.~\ref{overlapreformulated}. This feature is independent of the  auxiliary path followed, which characterizes the tremendous freedom at hand to reach the correct complex value.

The middle panel displays the resuts for $\varphi=\pi/2$. This case is of particular interest given that occupation numbers (i.e. $u^2_3=v^2_3=0.5$) have been chosen to ensure that $\langle \Phi | \breve{\Phi} \rangle=0$ at that angle. Whereas Eq.~\ref{overlapreformulated} was derived under the hypothesis that states along the manifold are not orthogonal to the initial state $| \Phi \rangle$, the fact that $| \breve{\Phi} \rangle = | \Phi(1) \rangle$ at the endpoint of the manifold is orthogonal to $| \Phi \rangle$ is gently obtained as the limit $\theta \rightarrow 1$, i.e. the corresponding result is recovered independently of the path followed. This key feature is analytically scrutinized in App.~\ref{gaugerotation}.

Let us now finally move to the lower panel corresponding to $\varphi=2\pi/3$, i.e. to a gauge angle that is larger than the value for which the overlap becomes zero along the manifold of gauge rotated states (full red line). In this case, the increment integration along the manifold looses the phase of the overlap as its norm goes through zero\footnote{To be in position to perform the calculation beyond $\varphi=\pi/2$ and make the figure, it is necessary to discretize the integral accross $\varphi=\pi/2$ in such a way that the principle value is obtained, thus bypassing the value $\varphi=\pi/2$ itself.}, i.e. the norm kernel acquires an extra minus sign. The extra minus sign reflects the invalidity of the exponential formula whenever a state along the auxiliary manifold is orthogonal to the initial state. This happens for $\varphi > \pi/2$ in the present context of gauge rotation whenever a conjugated pair $(k_1,\bar{k}_1)$ is characterized by $u_{k_1}=v_{k_1}$. 
More generally, the extrapolation of the exponential formula to the interval $\varphi \in [\pi/2,3\pi/2]$ differs from the correct value by a sign $(-1)^{p}$, where $p$ denotes the number of conjugated pairs characterized by $u_k=v_k=0.5$. This feature, analytically demonstrated in App.~\ref{gaugerotation}, has been checked numerically within the frame of the present toy model. Consistently, the sign becomes correct again when going through the next zero of the overlap, i.e. for $\varphi \in [3\pi/2,2\pi]$, independently of $p$.

The problem associated with going through zeros of the overlap along the auxiliary manifold appears at first sight as a limitation of the presently proposed method. One could rely on the analytical understanding of such a shortcoming to correct for the improper $(-1)^{p}$ sign. However, this procedure is not straightforwardly applicable to the overlap between arbitrary Bogoliubov states. Preferably, Fig.~\ref{results10levelsBCSmodel} illustrates that modifying the auxiliary manifold by multiplying  ${\cal W}$ or $\breve{{\cal W}}$ with a trivial Bogoliubov transformation ${\cal K}$ allows one to easily overcome this apparent limitation. Indeed, the complex value of the norm overlap is correctly captured at the price of generating such a random transformation ${\cal K}$\footnote{In a practical algorithm, one may want to generate a few random transformations ${\cal K}$ in order to improve the probability to find a safe path and reach convincingly the correct value.}. Given that the Onishi formula can be employed to anticipate whether or not zeros of the norm overlap occur along the auxiliary manifold, the potential problem can be easily identified and bypassed in any arbitrary situation.  

\subsection{Toy BCS states}
\label{resultstoyBCS}

\begin{figure}
\begin{center}
\includegraphics[width=6.5cm]{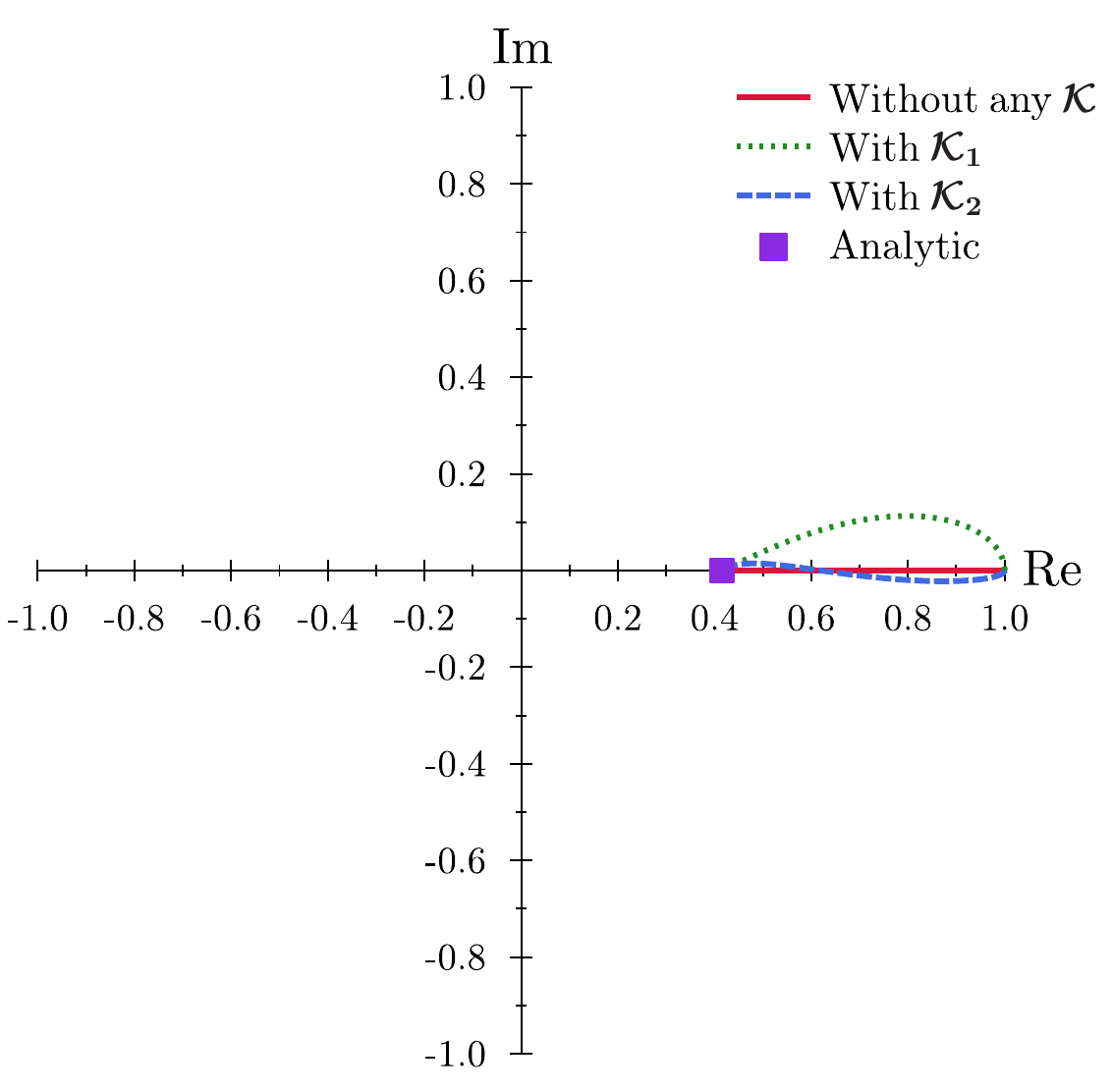}
\end{center}
\caption{(Color online) Norm overlap between BCS states $| \Phi \rangle$ and $| \breve{\Phi} \rangle$ represented in the complex plane. The purple square denotes the reference result obtained from Eq.~\ref{referenceoverlapBCStimesBCS}. The color lines correspond to the increment integration along the auxiliary manifolds linking $| \Phi \rangle$ to $| \breve{\Phi} \rangle$ obtained without (full red line) or with (dashed blue and dotted green lines) an additional trivial Bogoliubov transformation ${\cal K}$ (see Sec.~\ref{family}).}
\label{results10levelsBCSmodel2}
\end{figure}

The second example considered is even simpler than gauge rotation as it ensures that the overlap is strictly real. Keeping $| \Phi \rangle$ and $\bar{{\cal W}}$ as before, $| \breve{\Phi} \rangle$ is taken as a second generic BCS state. Consequently, the BCS transformation $\breve{\bar{{\cal W}}}$ is defined in the same single-particle basis by
\begin{subequations}
\label{BCStransforphibreve2}
\begin{align}
\breve{\bar{U}}(k,\bar{k}) &= 
\begin{pmatrix}
+\breve{u}_k & 0 \\
0 &  +\breve{u}_k
\end{pmatrix}  
\,\, , \\
\breve{\bar{V}}(k,\bar{k}) &=
\begin{pmatrix}
0 & +\breve{v}_k \\
-\breve{v}_k &  0
\end{pmatrix}  \,\, ,
\end{align}
\end{subequations}
with state $| \breve{\Phi} \rangle$ now reading as
\begin{align}
| \breve{\Phi} \rangle &\equiv  \prod_{k=1}^{5} (\breve{u}_k + \breve{v}_k c^{\dagger}_k c^{\dagger}_{\bar{k}}) | 0 \rangle \,\, ,
\end{align}
where $(\breve{u}_k, \breve{v}_k)$ are real and such that $\breve{u}^2_k + \breve{v}^2_k = 1$. The overlap between both states can be worked out straighforwardly
\begin{equation}
\frac{\langle \Phi | \breve{\Phi} \rangle}{\langle \Phi | \Phi \rangle}  = \prod_{k=1}^{5} (u_k \breve{u}_k + v_k \breve{v}_k)  \label{referenceoverlapBCStimesBCS}
\end{equation}
to provide its reference value. 

In Fig.~\ref{results10levelsBCSmodel2}, a numerical example is displayed\footnote{The actual value of the occupation numbers is irrelevant to the present proof-of-principle calculation and is thus not specified.}. The situation is similar to the case of gauge rotation, except that the overlap is real by construction. As a matter of fact, the manifold extracted straighforwardly from the BCS transformations given above does provide this real value by integrating incrementally over a path going along the real axis. Contrarily, performing an additional trivial complex Bogoliubov transformation ${\cal K}$ allows one to reach the real overlap by following a non-trivial path through the complex plane. In agreement with Eq.~\ref{zeroderivative}, and with all the other numerical illustrations presented in this work, the derivative of the norm overlap at $\theta =0$ is purely imaginary whenever $S^{00}\neq 0$ as is presently the case when applying a transformation ${\cal K}$.

\subsection{Toy Bogoliubov states}
\label{resultstoyBogo}

\begin{figure}
\begin{center}
\includegraphics[width=6.5cm]{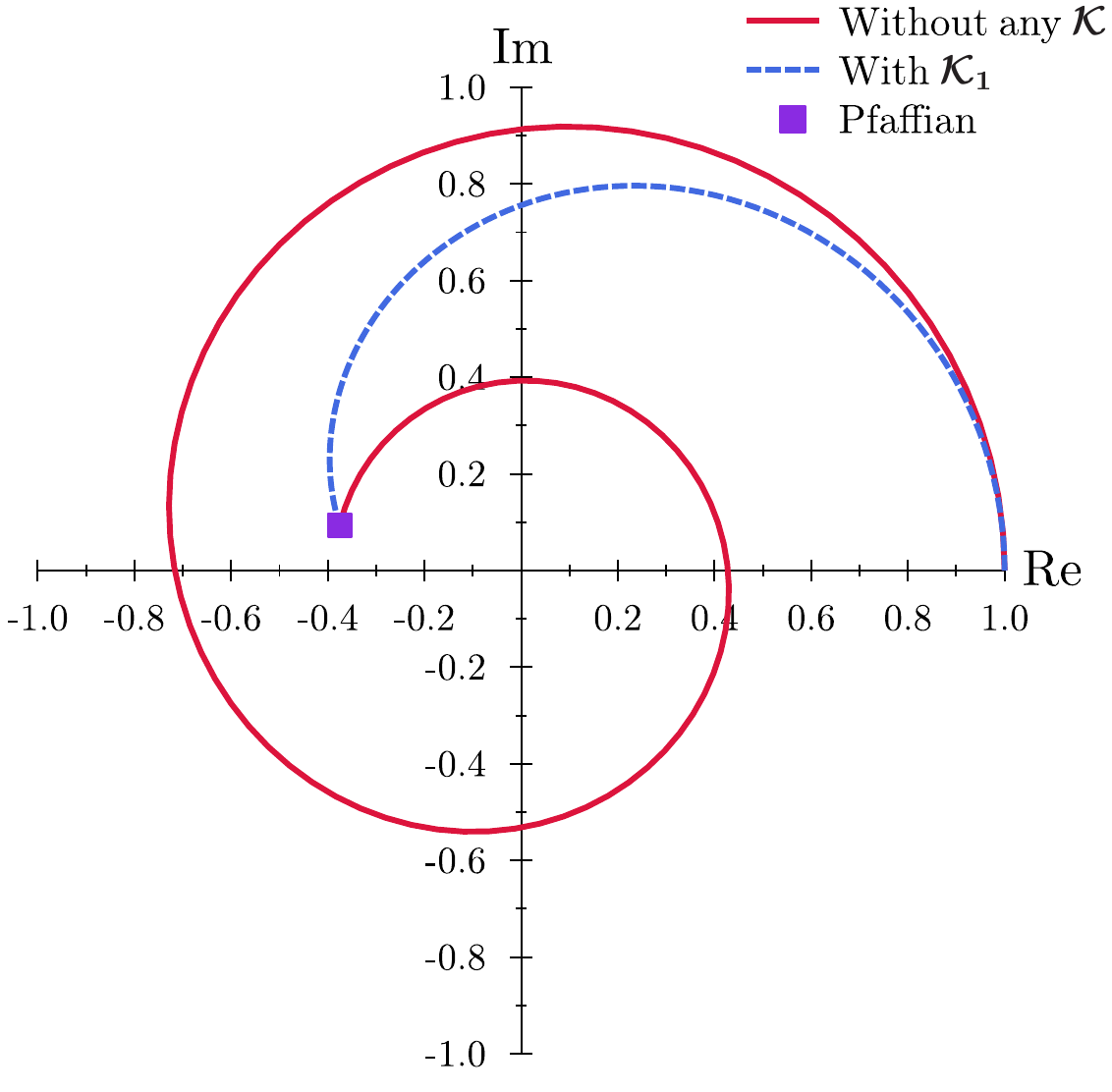}
\end{center}
\caption{(Color online) Norm overlap between Bogoliubov states $| \Phi \rangle$ and $| \breve{\Phi} \rangle$ represented in the complex plane. The purple square denotes the reference result obtained from Eq.~\ref{pfaffian}. The color lines correspond to the increment integration along the auxiliary manifolds linking $| \Phi \rangle$ to $| \breve{\Phi} \rangle$ obtained without (full red line) or with (dashed blue line) an additional trivial Bogoliubov transformation ${\cal K}$ (see Sec.~\ref{family}).}
\label{results10levelsBogomodel}
\end{figure}

It is now necessary to test the exponential formula for genuine Bogoliubov states. To do so, we start from the previous BCS toy model and construct more general Bogoliubov transformations of the form
\begin{subequations}
\begin{align}
{\cal W} &\equiv 
\begin{pmatrix}
L & 0 \\
0 &  L^{\ast}
\end{pmatrix}  
\begin{pmatrix}
\bar{U} & \bar{V}^{\ast} \\
\bar{V} &  \bar{U}^{\ast}
\end{pmatrix}  
  \,\, , \\
\breve{{\cal W}} &\equiv 
\begin{pmatrix}
\breve{L} & 0 \\
0 &  \breve{L}^{\ast}
\end{pmatrix}  
\begin{pmatrix}
\breve{\bar{U}} & \breve{\bar{V}}^{\ast} \\
\breve{\bar{V}} &  \breve{\bar{U}}^{\ast}
\end{pmatrix}  
\,\, ,
\end{align}
\end{subequations}
where $L$ and $\breve{L}$ are random complex $N\times N$ unitary matrices mimicking arbitrarily different canonical bases for the two Bogoliubov states. The Pfaffian approach to the norm overlap relies on the Thouless representation of the two states at play
\begin{subequations}
\begin{align}
| \Phi \rangle &\equiv \exp{\left(\frac{1}{2}\sum_{kk'} Z^{20}_{kk'} \, c^{\dagger}_k c^{\dagger}_{k'}\right)} | 0 \rangle  \,\, , \\
| \breve{\Phi} \rangle &\equiv \exp{\left(\frac{1}{2}\sum_{kk'} \breve{Z}^{20}_{kk'} \, c^{\dagger}_k c^{\dagger}_{k'}\right)} | 0 \rangle  \,\, ,
\end{align}
\end{subequations}
which is consistent with the phase convention $\text{Arg} (\langle 0 | \Phi \rangle) = \text{Arg} (\langle 0 | \breve{\Phi} \rangle)$. Eventually, the reference value of the norm overlap between both states is obtained as the Pfaffian of a skew-symmetric matrix~\cite{Robledo:2009yd,Robledo:2011ce,Avez:2011wr}
\begin{equation}
\frac{\langle \Phi | \breve{\Phi} \rangle}{\langle \Phi | \Phi \rangle}  = (-1)^{N(N+1)/2} \, \text{pf} 
\begin{pmatrix}
\breve{Z} & -1 \\
1 &  -Z^{\ast}
\end{pmatrix}   \, ,\label{pfaffian}
\end{equation}
which is presently computed using routines taken from Ref.~\cite{wimmer12a}.

In Fig.~\ref{results10levelsBogomodel}, the numerical example is displayed. The complex value of the overlap computed via the exponential formula matches the reference value obtained via Eq.~\ref{pfaffian}. This happens both without or with performing an additional trivial Bogoliubov transformation ${\cal K}$ (see Sec.~\ref{family}), i.e. the freedom associated with the auxiliary manifold remains fully operational here.

Let us remark that the algorithm works equally well if ${\cal W}$ and/or $\breve{{\cal W}}$ are characterized by fully occupied or fully empty paired canonical single-particle states. Additionnally, we have checked that the method works for odd-number parity states appropriate to the description of odd systems. Last but not least, and as was examplified in Sec.~\ref{resultsgaugerotation} for states differing by a gauge transformation of angle $\pi/2$, the method gently handles orthogonal states of identical number-parity.

\section{Many-body calculations}
\label{manybodymethods}

So far, the computation of the norm kernel associated with an arbitrary pair of states $\ket{\breve{\Phi}}$ and $\ket{\Phi}$ belonging to $\mathcal{M} \equiv \big\{ \ket{\Phi_{1}}, \ldots, \ket{\Phi_{N_{\text{set}}}}\big\}$ has been detailed. We now wish to discuss how this is to be done consistently in many-body calculations of interest that typically invoke the $N_{\text{set}}\times N_{\text{set}}$ hermitian matrix of uncorrelated norm overlaps
\begin{equation}
{\cal N}_{\mathcal{M}} \equiv \begin{pmatrix}
\langle \Phi_{1} | \Phi_{1} \rangle & \langle \Phi_{1} | \Phi_{2} \rangle & \cdots & & & \langle \Phi_{1} | \Phi_{N_{\text{set}}} \rangle \\
\langle \Phi_{2} | \Phi_{1} \rangle & \langle \Phi_{2} | \Phi_{2} \rangle &  & & & \\
\vdots &  & \ddots & & & \\
 &  &  & & & \\
\langle \Phi_{N_{\text{set}}} | \Phi_{1} \rangle &  &  & & & \langle \Phi_{N_{\text{set}}} | \Phi_{N_{\text{set}}} \rangle
\end{pmatrix} 
  \label{normmatrix} \, .
\end{equation}

\subsection{Generator coordinate method}
\label{resultstoyGCM}

We are first interested in discussing the situation typically encountered in generator coordinate method (GCM) calculations. In these calculations the full norm matrix associated with a set of pre-generated Bogoliubov vacua must be computed and diagonalized~\cite{ring80a,bender03b,Duguet:2013dga,Egido:2016bdz}. 

\subsubsection{Phase convention}

Given the perspective of computing a complete norm matrix as defined by Eq.~\ref{normmatrix}, the safest approach consists of taking the state fixing the phase convention within $\mathcal{M}$ itself. Consequently, we now choose $\ket{\bar{\Phi}}=\ket{\Phi_{1}}$ as such a pivot state, although any other state of the set would obviously be equally appropriate. This allows us to discussing a different phase convention from the one utilized in the above sections. This choice leads to expressing the phase convention stipulated in Sec.~\ref{discussion1} as
\begin{equation}
\text{Arg} (\langle \Phi_{1} | \Phi_{1} \rangle) = \text{Arg} (\langle \Phi_{1} | \Phi_{2} \rangle) \ldots = \text{Arg} (\langle \Phi_{1} | \Phi_{N_{\text{set}}} \rangle) = 0 \, , \label{phaseconventionMRset}
\end{equation}
given that $\langle \Phi_{1} | \Phi_{1} \rangle$ is real.

\subsubsection{Procedure}

The computation of the $N_{\text{set}}(N_{\text{set}}+1)/2$ independent overlaps making up the norm matrix follows three successive steps
\begin{enumerate}
\item The $N_{\text{set}}$ diagonal elements are trivially obtained by normalizing the members of the set, i.e. by imposing that $\langle \Phi_{l} | \Phi_{l} \rangle=1$ for $l=1, \ldots, N_{\text{set}}$. 
\item The $N_{\text{set}}-1$ remaining elements of the first row are computed by introducing the  $N_{\text{set}}-1$ operators 
\begin{eqnarray}
S[l] &\equiv& S^{00}[l]_1 + \frac12 \Tr\left(S^{11}[l]_1\right) \label{eq:1bodynobogok} \\
&+& \frac12  
\left(\,\beta^{\dagger}[1] \hspace{0.2cm} \beta[1] \,\right)
\begin{pmatrix}
S^{11}[l]_1 & S^{20}[l]_1 \\
-S^{02}[l]_1 &  -S^{11 \ast}[l]_1
\end{pmatrix} 
\begin{pmatrix}
\beta[1] \\
\beta^{\dagger}[1]
\end{pmatrix}  \, , \nonumber
\end{eqnarray}
with $l=2, \ldots, N_{\text{set}}$, such that
\begin{equation}
| \Phi_l \rangle   \equiv e^{iS[l]} |  \Phi_1 \rangle \, . \label{defphil}
\end{equation}
The operator $S[l]$ depends {\it implicitly} on the pivot state $\ket{\Phi_{1}}$ given that it connects $| \Phi_l \rangle$ to it. In Eq.~\ref{eq:1bodynobogok} the operator has been represented in the quasiparticle basis $\{\beta_k[1];\beta^{\dagger}_k[1]\}$ associated with the pivot state such that matrices $S^{ij}[l]_1$  {\it explicitly} depend on the index, i.e. $1$ here, labelling the quasi-particle basis used. Of course, the operator $S[l]$ can be equally represented in the quasi-particle basis associated with any other state $\ket{\Phi_{m}}$ of the set, in which cases the associated matrices are denoted as $S^{ij}[l]_m$. 

With these definitions at hand, the hermitian matrix 
\begin{equation}
{\cal S}[l]_1 \equiv \begin{pmatrix}
S^{11}[l]_1 & S^{20}[l]_1 \\
-S^{02}[l]_1 &  -S^{11 \ast}[l]_1
\end{pmatrix} 
\end{equation}
entering $S[l]$ must be extracted according to the procedure outlined in Sec.~\ref{formalismkernel} (and summarized in Sec.~\ref{algorithm}). From there, the $N_{\text{set}}-1$ off-diagonal norm overlaps can be computed by applying Eq.~\ref{eq:genov} for $\langle \Theta | \equiv \langle \Phi_1 |$ and $\theta=1$ 
\begin{equation}
\label{eq:genovHFBNO1l_A}
\begin{split}
\frac{\langle \Phi_1 | \Phi_l \rangle}{\langle \Phi_1 | \Phi_1 \rangle} &=  e^{i S^{00}[l]_1} e^{ \frac{i}{2} \int_0^1 d\theta \Tr \left( S^{02}[l]_1 R^{--}[\langle \Phi_1 |,| \Phi_{l1}(\theta) \rangle] \right)} \, ,
\end{split}
\end{equation}
where the elementary contractions
\begin{equation}
R^{--}_{k_1k_2}[\langle \Phi_1 |,| \Phi_{l1}(\theta) \rangle] \equiv \frac{\langle \Phi_1 | \beta^{\phantom{\dagger}}_{k_1}[1]\beta^{\phantom{\dagger}}_{k_2}[1] | \Phi_{l1}(\theta) \rangle}{\langle \Phi_1 | \Phi_{l1}(\theta) \rangle} 
\end{equation}
run over the manifold 
\begin{equation}
\mathcal{M}[| \Phi_1 \rangle,S[l]] \equiv \big\{ \ket{\Phi_{l1}(\theta)} \equiv e^{i \theta S[l]}  \ket{\Phi_1} \, , \, \theta \in [0, 1] \big\} \nonumber
\end{equation}
connecting $| \Phi_1 \rangle$ to $| \Phi_l \rangle$. The remaining unkwown $S^{00}[l]_1$ entering $S[l]$ is fixed by enforcing Eq.~\ref{phaseconventionMRset}, which translates into
\begin{equation}
S^{00}[l]_1 = - \Re e \frac12 \int_0^1 d\theta \, \Tr\left( S^{02}[l]_1 R^{--}[\langle \Phi_1 |,| \Phi_{l1}(\theta) \rangle] \right)\, ,
\end{equation}
making the $N_{\text{set}}-1$ overlaps real and reading as
\begin{equation}
\label{eq:genovHFBNO1l_B}
\begin{split}
\frac{\langle \Phi_1 | \Phi_l \rangle}{\langle \Phi_1 | \Phi_1 \rangle} &=  e^{ -\Im m \frac{1}{2} \int_0^1 d\theta \Tr \left( S^{02}[l]_1 R^{--}[\langle \Phi_1 |,| \Phi_{l1}(\theta) \rangle] \right)} \, .
\end{split}
\end{equation}
\item With the $N_{\text{set}}-1$ operators $S[l]$ at hand, the remaining $(N_{\text{set}}-1)(N_{\text{set}}-2)/2$ independent norm kernels can be calculated consistently. Starting from Eq.~\ref{defphil} and applying Eq.~\ref{eq:genov} for $\langle \Theta | \equiv \langle \Phi_m |$ and $\theta=1$ leads, for $1<m<l\leq N_{\text{set}}$, to
\begin{equation}
\label{eq:genovHFBNOml}
\begin{split}
\frac{\langle \Phi_m | \Phi_l \rangle}{\langle \Phi_m | \Phi_1 \rangle} &=  e^{i S^{00}[l]_m} e^{ \frac{i}{2} \int_0^1 d\theta \Tr \left( S^{02}[l]_m R^{--}[\langle \Phi_m |,| \Phi_{l1}(\theta) \rangle] \right)} \, ,
\end{split}
\end{equation}
where matrices $S^{ij}[l]_m$ result from expressing $S[l]$ in the quasiparticle basis associated with $| \Phi_m \rangle$ and where the elementary contractions
\begin{equation}
R^{--}_{k_1k_2}[\langle \Phi_m |,| \Phi_{l1}(\theta) \rangle] \equiv \frac{\langle \Phi_m | \beta^{\phantom{\dagger}}_{k_1}[m]\beta^{\phantom{\dagger}}_{k_2}[m] | \Phi_{l1}(\theta) \rangle}{\langle \Phi_m | \Phi_{l1}(\theta) \rangle} 
\end{equation}
run over the manifold $\mathcal{M}[| \Phi_1 \rangle,S[l]]$. Similarly to before, these elementary contractions are easily computed from the Bogoliubov transformation linking the bra and the ket, itself being obtained from the Bogoliubov transformations associated with $| \Phi_{1}\rangle$, $| \Phi_{l}\rangle$ and $| \Phi_{m}\rangle$. Since $\langle \Phi_m | \Phi_1 \rangle$ is among the $N_{\text{set}}-1$ overlaps already computed in step 2 (Eq.~\ref{eq:genovHFBNO1l_B}), Eq.~\ref{eq:genovHFBNOml} completes the norm matrix. While the diagonal, the first row and the first column are real by virtue of choosing $| \Phi_1 \rangle$ as the pivot state to fix the phase, the remaining entries of the norm matrix are a priori complex.
\end{enumerate}

\subsubsection{Numerical application}

We extend the toy calculation of Sec.~\ref{resultstoyBogo} to a set $\mathcal{M} \equiv \big\{ | \Phi_1 \rangle, | \Phi_2 \rangle, | \Phi_3 \rangle\big\}$ of three different Bogoliubov states and compute their associated norm matrix 
\begin{equation}
{\cal N}_{\mathcal{M}} \equiv 
\begin{pmatrix}
\langle \Phi_{1} | \Phi_{1} \rangle & \langle \Phi_{1} | \Phi_{2} \rangle & \langle \Phi_{1} | \Phi_{3} \rangle \\
\langle \Phi_{2} | \Phi_{1} \rangle & \langle \Phi_{2} | \Phi_{2} \rangle & \langle \Phi_{2} | \Phi_{3} \rangle \\
\langle \Phi_{3} | \Phi_{1} \rangle & \langle \Phi_{3} | \Phi_{2} \rangle & \langle \Phi_{3} | \Phi_{3} \rangle 
\end{pmatrix} 
  \label{normmatrixGCM3states}
\end{equation}
according to the algorithm detailed above. Doing so, we have switched from the phase convention  associated with choosing the pivot state as $| \bar{\Phi} \rangle = | 0 \rangle$ to choosing it within the MR set, i.e. $| \bar{\Phi} \rangle = | \Phi_1 \rangle$ here. Given that the Pfaffian method used to benchmark our method explicitly relies on the first choice, the present example is meant to underline that what matters is not the calculation of a given norm overlap per se but the production of a consistent norm matrix, i.e. a norm matrix whose eigenvalues are insensitive to the state chosen to fix the phase convention.

\begin{figure}
\begin{center}
\includegraphics[width=8.5cm]{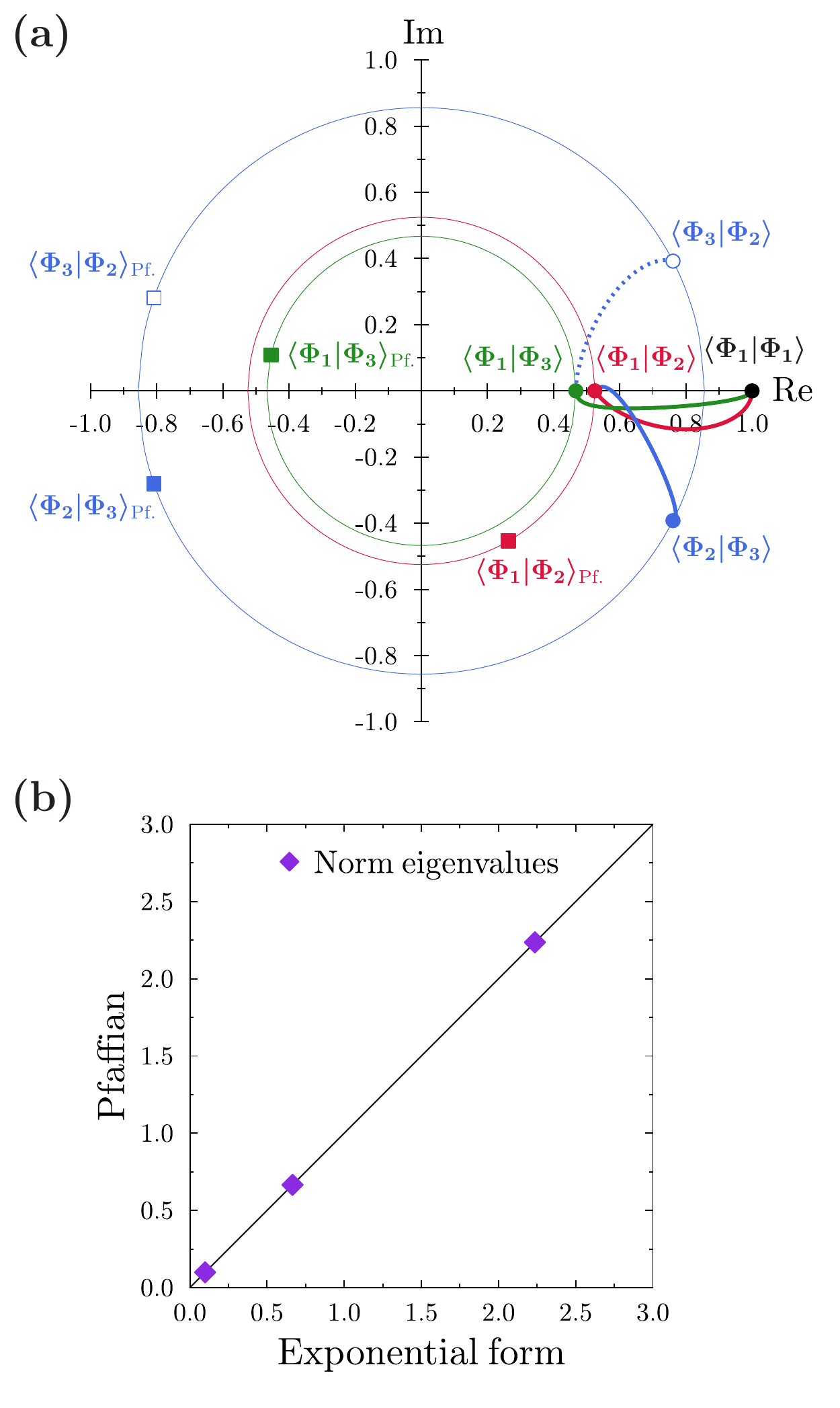}
\end{center}
\caption{(Color online) Upper panel: norm kernels making up the uncorrelated norm matrix associated with the MR set of three normalized Bogoliubov states $\{| \Phi_1 \rangle, | \Phi_2 \rangle, | \Phi_3 \rangle\}$. Squares denote values obtained from the Pfaffian method whereas circles denote those obtained from the present scheme using $| \Phi_1 \rangle$ as the pivot state for the phase convention. The lines provide the auxiliary pathes followed from one overlap to the other, starting from $\langle \Phi_1 | \Phi_1 \rangle = 1$. Lower panel: eigenvalues of the norm matrix obtained on the basis of the Pfaffian method against those obtained with the exponential formula. }
\label{results10levelsGCMBogomodel}
\end{figure}

Figure~\ref{results10levelsGCMBogomodel} displays the results of the numerical example. The upper panel shows individual norm overlaps making up the norm matrix. Squares represent the values obtained from the Pfaffian method whereas circles denote those obtained from the present method. Thick lines provide the auxiliary pathes followed from one overlap to the other, starting from $\langle \Phi_1 | \Phi_1 \rangle = 1$. Consistently with the scheme exposed above, all the overlaps involving the pivot state $| \Phi_1 \rangle$ are real. Furthermore, the complex conjugate values $\langle \Phi_2 | \Phi_3 \rangle$ and $\langle \Phi_3 | \Phi_2 \rangle$ are consistently obtained by choosing to go through $| \Phi_2 \rangle$ or $| \Phi_3 \rangle$ first. Last but not least, three circles help visualize that, while individual kernels differ in both methods, they only do so by a complex phase. Eventually, the lower panel demonstrates that the eigenvalues of the norm matrices obtained from both methods are identical, thus showing the consistency of both calculations and the independence on the phase convention used.

\subsection{Symmetry restoration}
\label{symmetryrestoration}

To complete the discussion, we now move to symmetry restoration calculations~\cite{ring80a,bender03b,Duguet:2013dga,Egido:2016bdz}. Focusing on particle number restoration associated with global gauge symmetry as an example, we wish to illustrate how the two phase conventions used earlier can be equally employed on the basis of projection or diagonalization methods. The present section relies on useful details regarding global gauge transformation and particle number restoration given in Apps.~\ref{gaugerotation} and~\ref{appendixPNR}, respectively.

\subsubsection{U(1) group and associated MR set}

The particle-number restoration relates to the one-parameter Abelian Lie group $U(1)$ defined by
\begin{equation}
U(1) \equiv \left\{ R(\varphi) \equiv e^{i \varphi A}  \equiv e^{iS(\varphi)}, \,  \varphi \in [0,2\pi] \right\} \,  , \label{U1group}
\end{equation}
where $\varphi$ denotes the gauge angle and $R(\varphi)$ is a unitary representation of global gauge transformations over Fock space.  The particle-number operator $A$ is a generator of the group with
\begin{equation}
A\equiv -i \frac{d}{d\varphi} \, .
\end{equation}
The irreducible representations (IRREPs) of the group read as
\begin{align}
 \label{eq:pnrirrep}
\langle \Psi^{\text{A}} | R(\varphi) | \Psi^{\text{A}'} \rangle &\equiv   e^{i \varphi \text{A}} \, \delta_{\text{A}\text{A}'} \, ,
\end{align}   
where $| \Psi^{\text{A}} \rangle$  is a normalized eigenstate of the particle number operator $A$ associated with eigenvalue $\text{A}$. The orthogonality of the IRREPs is expressed as
\begin{equation}
 \label{orthoirr}
  \int_{0}^{2\pi} d\varphi \, e^{-i \varphi \text{A}} \, e^{i \varphi \text{A}'} = 2\pi \, \delta_{\text{A}\text{A}'}  \, .
\end{equation}      

Given a Bogoliubov state $| \Phi \rangle$ breaking $U(1)$ symmetry, the MR set to be considered to restore good particle number is given by the orbit of the group
\begin{align}
\mathcal{M}_{U(1)} &\equiv \left\{ | \Phi(\varphi) \rangle \equiv R(\varphi) | \Phi \rangle , \,  \varphi \in [0,2\pi] \right\} \, , 
\end{align}
with $| \Phi(0) \rangle \equiv | \Phi \rangle$ by definition. 

\subsubsection{Norm matrix}

The two available techniques to restore good particle number make partial or full use of the norm matrix ${\cal N}_{\mathcal{M}_{U(1)}}$ generated out of the MR set $\mathcal{M}_{U(1)}$; see App.~\ref{appendixPNR} for details. In practical calculations, the gauge angle $\varphi$ is discretized by appropriately selecting $\{\varphi_i \in [0,2\pi], \, i=1,\ldots, N_{\text{set}}\}$, where $\varphi_1 =0$ by convention. As a result, the norm matrix ${\cal N}_{\mathcal{M}_{U(1)}}$ at play takes the form
\begin{equation}
\begin{pmatrix}
\langle \Phi(0) | \Phi(0) \rangle & \langle \Phi(0) | \Phi(\varphi_{2}) \rangle & \cdots &  \langle \Phi(0) | \Phi(\varphi_{N_{\text{set}}}) \rangle \\
\langle \Phi(\varphi_{2}) | \Phi(0) \rangle & \langle \Phi(\varphi_{2}) | \Phi(\varphi_{2}) \rangle &  &   \\
\vdots &  & \ddots &   \\
\langle \Phi(\varphi_{N_{\text{set}}}) | \Phi(0) \rangle &  &  &   \langle \Phi(\varphi_{N_{\text{set}}}) | \Phi(\varphi_{N_{\text{set}}}) \rangle
\end{pmatrix} 
  \label{normmatrixPNR} \nonumber
\end{equation}

The unitary representation of $U(1)$ introduced in Eq.~\ref{U1group} amounts to choosing the set of hermitian operators $S(\varphi)$ connecting the unrotated state $| \Phi \rangle$ with the rotated ones $| \Phi(\varphi) \rangle$ according to $S(\varphi) \equiv \varphi A$, i.e. 
\begin{subequations}
\label{transfomatricesssymgauge}
\begin{align}
s^{00}(\varphi) &\equiv 0 \, , \label{transfomatricesssymgauge1} \\
s^{11}(\varphi) &\equiv \varphi A \, , \label{transfomatricesssymgauge2} \\
s^{20}(\varphi) &\equiv 0 \, , \label{transfomatricesssymgauge3} \\
s^{02}(\varphi) &\equiv 0  \, , \label{transfomatricesssymgauge4}
\end{align}
\end{subequations}
or equivalently
\begin{subequations}
\label{transfomatricesSsymgaugecorps}
\begin{align}
S^{00}(\varphi) &\equiv \varphi A^{00} = \varphi \, \text{Tr}(V^{\ast}V^T) \, , \label{transfomatricesSsymgaugecorps1} \\
S^{11}(\varphi) &\equiv \varphi A^{11} = \varphi \left[U^\dagger U - V^\dagger V\right] \, , \label{transfomatricesSsymgaugecorps2} \\
S^{20}(\varphi) &\equiv \varphi A^{20} = \varphi \left[U^\dagger V^* - V^\dagger U^*\right] \, , \label{transfomatricesSsymgaugecorps3} \\
S^{02}(\varphi) &\equiv \varphi A^{02} = \varphi \left[U^T  V - V^T  U \right]  \, . \label{transfomatricesSsymgaugecorps4}
\end{align}
\end{subequations}
This representation amounts to employing the phase convention
\begin{equation}
\text{Arg} (\langle 0 | \Phi(\varphi) \rangle) = \text{Arg} (\langle 0 | \Phi \rangle) \, , \, \forall \, \varphi \in [0,2\pi] \, ,\label{standardPC}
\end{equation}
i.e. to taking\footnote{We assume here a fully paired even number-parity Bogoliubov state that is not orthogonal to the particle vacuum. More generally, the appropriate state of reference is the Slater determinant $| \bar{\Phi} \rangle$ built out of the $N_{\text{occ}}$ canonical single-particle states that are fully occupied in $| \Phi \rangle$. In this case, the standard constant phase convention~\ref{standardPC} with respect to the particle vacuum is to be replaced by
\begin{equation}
 \frac{\langle \bar{\Phi} | \Phi(\varphi) \rangle}{\langle \bar{\Phi} | \Phi \rangle} = e^{+i \varphi N_{\text{occ}}} \, .
\end{equation}} 
$| \bar{\Phi} \rangle \equiv | 0 \rangle$. This choice constitutes the standard phase convention used in symmetry restoration calculations. 

In this situation, and as demonstrated in App.~\ref{gaugerotation} on the basis of the general development of Sec.~\ref{formalismkernel}, the overlaps appearing on the first row of the norm matrix ${\cal N}_{\mathcal{M}_{U(1)}}$ are given by
\begin{equation}
\frac{\langle \Phi | \Phi(\varphi) \rangle}{\langle \Phi | \Phi \rangle}  = e^{i \varphi \, A^{00}} e^{\frac{i}{2} \int_0^{\varphi} d\phi  \Tr \left(A^{02} R^{--}(\phi)\right)} \, , \label{normoverlapgaugebulk}
\end{equation}
where $R^{--}(\phi)$ is defined in  Eq.~\ref{eq:rmmsecond2}. Following similar steps, the norm matrix is completed with entries
\begin{align}
\frac{\langle \Phi(\varphi') | \Phi(\varphi) \rangle}{\langle \Phi(\varphi') | \Phi(\varphi') \rangle}  &= \frac{\langle \Phi(\varphi') | \Phi(\varphi) \rangle}{\langle \Phi | \Phi \rangle} \label{normoverlapgaugebulk2} \\
&= \frac{\langle \Phi | \Phi(\varphi-\varphi') \rangle}{\langle \Phi | \Phi \rangle} \nonumber \\
&= e^{i (\varphi-\varphi') \, A^{00}} e^{\frac{i}{2} \int_0^{\varphi-\varphi'} d\phi  \Tr \left(A^{02} R^{--}(\phi)\right)} \, . \nonumber 
\end{align}
While the projection technique only makes use of the first row of the norm matrix, the diagonalization method exploit the norm matrix in full; see App.~\ref{appendixPNR} for details.

\subsubsection{Alternative phase convention}

The overlaps appearing on the first row of the norm matrix are not real when given under the form of Eq.~\ref{normoverlapgaugebulk}. This result is consistent with the fact that the phase convention used is different from the one advocated in connection with the GCM calculation described in Sec.~\ref{resultstoyGCM} above. To relate both conventions in the present context, we introduce the 
operator\footnote{For the sake of simplicity, we do not consider here additional trivial transformations $\mathcal{K}$ to the Bogoliubov transformations associated with the bra or the ket (see Sec.~\ref{family} for details). In that case the operator $S(\varphi)$ to be extracted cannot be expressed simply in terms of the operator $A$. Still, the phase difference 
$\tilde{s}^{00}(\varphi)$ associated with both phase conventions can be worked out in a similar way as here.}
\begin{align}
 \label{eq:pnrrep2}
\tilde{S}(\varphi) &\equiv \tilde{s}^{00}(\varphi) + S(\varphi) \nonumber \\  
&=  \tilde{s}^{00}(\varphi) +  \varphi A \, , 
\end{align}      
where the (gauge-angle dependent) constant is given by
\begin{equation}
 \label{constant}
\tilde{s}^{00}(\varphi) \equiv -\varphi \, A^{00} -  \Re e \frac{1}{2} \int_0^{\varphi} d\phi  \Tr \left(A^{02} R^{--}(\phi)\right) \, .
\end{equation}     
Obviously, $\tilde{S}(\varphi)$ and $S(\varphi)$ are identical in any basis representation except for a constant. With this definition at hand we introduce the modified manifold of gauge-rotated states according to 
\begin{equation}
\tilde{\mathcal{M}}_{U(1)} \equiv \left\{ | \tilde{\Phi}(\varphi) \rangle \equiv e^{i\tilde{S}(\varphi)} | \Phi \rangle  ,  \varphi \in [0,2\pi] \right\} \, , \nonumber
\end{equation}
such that $| \tilde{\Phi}(0) \rangle = | \Phi \rangle$ and $| \tilde{\Phi}(\varphi) \rangle= e^{i \tilde{s}^{00}(\varphi)} | \Phi(\varphi) \rangle$. 

The overlaps appearing on the first row of the modified norm matrix ${\cal N}_{\tilde{\mathcal{M}}_{U(1)}}$ are equal to 
\begin{align}
\frac{\langle \Phi | \tilde{\Phi}(\varphi) \rangle}{\langle \Phi | \Phi \rangle}  &= e^{- \Im m \frac{1}{2} \int_0^{\varphi} d\phi  \Tr \left(A^{02} R^{--}(\phi)\right)} \nonumber \\
&= e^{i\tilde{s}^{00}(\varphi)} \frac{\langle \Phi | \Phi(\varphi) \rangle}{\langle \Phi | \Phi \rangle} \, , \label{normoverlapgaugebulk3}
\end{align}
and are real in agreement with the phase convention relying on the pivot state $| \bar{\Phi} \rangle=| \Phi \rangle$. Following the procedure detailed in Sec.~\ref{resultstoyGCM}, the modified norm matrix is completed thanks to the complex entries
\begin{align}
\langle \tilde{\Phi}(\varphi') | \tilde{\Phi}(\varphi) \rangle  &=  e^{i\left(\tilde{s}^{00}(\varphi)-\tilde{s}^{00}(\varphi')\right)} \langle \Phi(\varphi') | \Phi(\varphi) \rangle \label{normoverlapgaugebulk4} \\
&\neq  \langle \Phi | \tilde{\Phi}(\varphi-\varphi') \rangle \, , \nonumber 
\end{align}
the last inequality being due to the fact that\footnote{This property forbids the set of transformations $\{e^{i\tilde{S}(\varphi)}  ,  \varphi \in [0,2\pi]\}$ to constitute a representation of the $U(1)$ group.}
\begin{equation}
\tilde{s}^{00}(\varphi)-\tilde{s}^{00}(\varphi')\neq \tilde{s}^{00}(\varphi-\varphi') \, .
\end{equation}
Effectively, the change of phase convention corresponds to unitarily transforming the norm matrix according to 
\begin{equation}
{\cal N}_{\tilde{\mathcal{M}}_{U(1)}} = Q^{\dagger} {\cal N}_{\mathcal{M}_{U(1)}} Q
\end{equation}
where 
\begin{equation}
Q \equiv 
\begin{pmatrix}
1 & 0 & \cdots &  0 \\
0 & e^{-i \tilde{s}^{00}(\varphi_{2})}  &  &  \vdots \\
\vdots &  & \ddots &   \\
0 & \cdots &  &   e^{-i \tilde{s}^{00}(\varphi_{N_{\text{set}}})} 
\end{pmatrix} 
  \label{matricepassage} \, .
\end{equation}
The Hamiltonian matrix ${\cal H}_{\mathcal{M}_{U(1)}}$ is transformed accordingly such that the eigenvalues of both matrices are independent of the phase convention. This key feature is illustrated numerically below. As was implicitly clear in Sec.~\ref{resultstoyGCM}, modifying the phase convention in GCM calculations also corresponds to unitarily transforming the norm and Hamiltonian matrices. 

From the point of view of using a projector, the change of phase convention corresponds to using
\begin{align}
| \Phi^{\text{A}} \rangle &\equiv \int_{0}^{2\pi} \frac{d\varphi}{2\pi} \, e^{-i \left(\tilde{s}^{00}(\varphi)+\varphi \text{A}\right)}  \, | \tilde{\Phi}(\varphi) \rangle \, , \label{projectionmodified}
\end{align}
instead of Eq.~\ref{projection}. Consequently, the particle-number restored energy becomes
\begin{subequations}
\begin{align}
\text{E}^{\text{A}} &\equiv \frac{\langle \Phi^{\text{A}}  | H | \Phi^{\text{A}}  \rangle}{\langle \Phi^{\text{A}}  | \Phi^{\text{A}}  \rangle} \\
&=  \frac{\int_{0}^{2\pi} \!d\varphi \,  e^{-i \left(\tilde{s}^{00}(\varphi)+\varphi \text{A}\right)}  \, \, \langle \Phi | H | \tilde{\Phi}(\varphi) \rangle}{\int_{0}^{2\pi} \!d\varphi \,  e^{-i \left(\tilde{s}^{00}(\varphi)+\varphi \text{A}\right)}  \,\, \langle \Phi  | \tilde{\Phi}(\varphi) \rangle}  \, ,
\end{align}
\end{subequations}
instead of Eq.~\ref{projEannexe}. Eventually, the two phase conventions can be equally applied such that observables do not depend on this choice. In the projection technique, however, moving away from the phase convention associated with the standard representation of the $U(1)$ groups comes with the price of explicitly compensating the modified phase of gauge-rotated states by multiplying its irreducible representations accordingly.

\subsubsection{Numerical application}

We now illustrate numerically the equal validity of the projection and the diagonalization methods along with the freedom regarding the phase convention used. To do so, we employ the  toy model of Sec.~\ref{resultsgaugerotation}. 

\begin{figure}[t!]
\centering  
  \includegraphics[height=7.5cm]{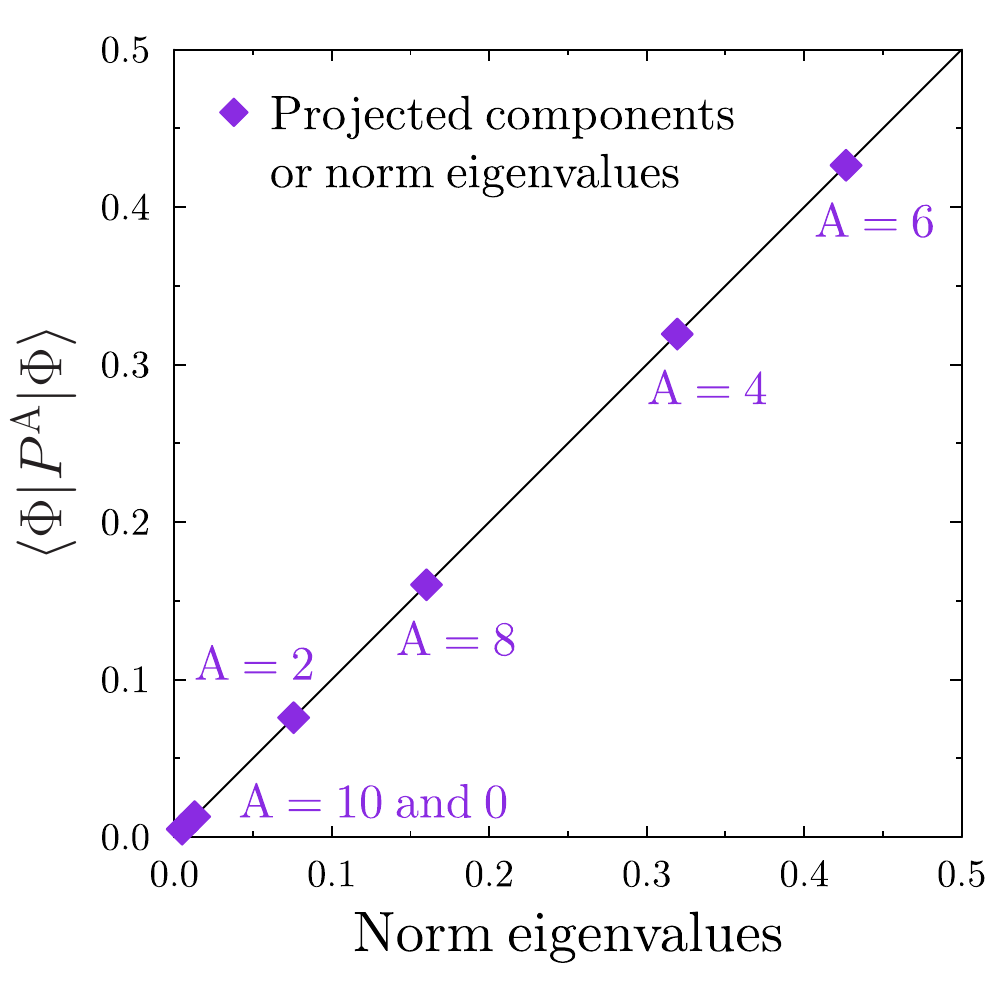}
\caption{
\label{fig:pvsd}
(Color online)
Components $\langle \Phi | P^{\text{A}} |  \Phi \rangle$ of the normalized projected states  $| \Lambda^{\text{A}} \rangle$ in the BCS state of reference  $| \Phi \rangle$ against eigenvalues of the corresponding norm matrix ${\cal N}_{\tilde{\mathcal{M}}_{U(1)}}$.
}
\end{figure}

In Fig.~\ref{fig:pvsd}, we compare the components $\langle \Phi | P^{\text{A}} |  \Phi \rangle$ of the normalized projected states in the BCS state of reference with the eigenvalues $n_{\text{A}}$ of the norm matrix ${\cal N}_{\tilde{\mathcal{M}}_{U(1)}}$, the latter making use of the phase convention associated with the pivot state $| \bar{\Phi} \rangle = | \Phi \rangle$.  The results of both methods match perfectly. This proves, once again, the sole necessity to choose a consistent phase convention within the set. Using 6 points to discretize the gauge angle, the values $n_{\text{A}}$ are all perfectly converged independently of the method used. Using less points first degrades the smallest eigenvalues.

\subsection{Mixed phase conventions}

The large freedom associated with (appropriate) phase conventions has been illustrated both for GCM and symmetry restoration calculations. It must eventually be made clear that mixed phase conventions can also be employed. While using a state belonging to the MR set as the pivot state is highly appropriate for GCM-type mixing, the phase convention implicitly associated with the standard unitary representations of symmetry groups must be favored for states relating to one another via symmetry transformations. Consequently, a mixed convention can typically be used to fix the phases within the MR set of interest on the basis of the following algorithm
\begin{enumerate}
\item Fix the phase associated with the subset of states that do not relate to another via symmetry transformations by using the pivot state within this subset.
\item For each of these states, generate the manifold of states related to it via symmetry transformations on the basis of standard unitary representations of associated symmetry groups. The phase of the states belonging to this sub-manifold is typically fixed with respect to a state, e.g. $| 0 \rangle$, that is outside of it.
\end{enumerate}
Following the developments of the preceeding sections, the complete norm matrix can be built incrementally on the basis of this mixed phase convention.

\section{Conclusions}
\label{Conclusions}

The present paper proposes an efficient and versatile method to compute the overlap between {\it arbitrary} Bogoliubov states $| \Phi \rangle$ and $| \breve{\Phi} \rangle$. The associated formula reads as the exponential of an integral of the off-diagonal kernel of an operator $S$ along an auxiliary manifold of Bogoliubov states linking $| \Phi \rangle$ and $| \breve{\Phi} \rangle$. The operator $S$, which is to be extracted during the procedure, is the generator of a unitary transformation linking both Bogoliubov vacua. All in all, the norm overlap between arbitrary Bogoliubov states is computed, without any phase ambiguity, via elementary linear algebra operations. The method can be used in any configuration mixing of orthogonal and non-orthogonal product states.

We have performed several numerical illustrations based on toy models of increasing complexity. When dealing with general Bogoliubov states that do not share a common discrete symmetry, such as simplex or time reversal, the results  are benchmarked against the Pfaffian method~\cite{Robledo:2009yd}. The versatility of the method allows one to reach the correct complex value by integrating over pathes associated with many different auxiliary manifolds. This is convenient, in particular, to bypass potential zeros of the overlap along the manifold.

The method is already interesting in itself, even though the efficient Pfaffian alternative already exists. In particular, the  natural and intuitive closed-form expression sheds a new light on the intimate content of the overlap between arbitrary Bogoliubov states. More importantly, the method is appealing from the point of view of its generic character and of the possible extensions it offers. In particular, it allows one to address {\it correlated} norm kernels at play in recently developped particle-number-restored Bogoliubov coupled-cluster (PNR-BCC) and particle-number-restored many-body perturbation  (PNR-BMBPT) ab initio theories~\cite{Duguet:2015yle}. It is the aim of a forthcoming paper~\cite{arthuis17a} to extend the present work to such general norm kernels from which {\it uncorrelated} kernels, i.e. straight overlaps between Bogoliubov vacua, are recovered as a particular case. Furthermore, the generality of the method makes possible to envision computing overlaps based on other many-body states than Bogoliubov product states.

\begin{acknowledgments}
The authors thank J.-P. Ebran for motivating them to engage in this study. The authors further thank P. Arthuis, M. Bender and V. Som\`a for proofreading the manuscript as well as M. Bender and M. Drissi for fruitful discussions.
\end{acknowledgments}

\begin{appendix}

\section{Unicity of ${\cal S}$}
\label{unicityS}

We illustrate why the matrix ${\cal S}$ must be computed from the {\it principal} logarithm of ${\cal X}$ and why it is, as such, uniquely defined. To achieve this goal, we use a highly schematic model in which the matrix ${\cal X}$ is two dimensional and diagonal. This is sufficient to make the point and necessary to do so transparently. 

As ${\cal X}={\cal X}(1)$ is a proper unitary Bogoliubov transformation, it displays the typical form
\begin{equation}
{\cal X}(1) =
\begin{pmatrix}
e^{-is} & 0 \\
0 &  e^{+is}
\end{pmatrix}  \, , 
\end{equation}
with $s \in ]-\pi,\pi]$ without any loss of generality. Based on Eq.~\ref{exprepresentationWtheta} indicating that ${\cal X}(\theta)=\exp{(-i\theta{\cal S})}$, our goal is to extract ${\cal S}$ at a particular value of $\theta$ before generating the entire family of transformations ${\cal X}(\theta)$. Knowing ${\cal X}(1)$, we set $\theta=1$ to extract ${\cal S}$ according to
\begin{equation}
{\cal S}^{(1)}  = i \log {\cal X}(1) \, , \label{extraction1st}
\end{equation}
where the superscript underlines that ${\cal S}$ is extracted from Eq.~\ref{exprepresentationWtheta} at the specific value $\theta=1$. Authorizing the use of a non-principal logarithm, ${\cal S}^{(1)}$ reads as
\begin{equation}
{\cal S}^{(1)} =
\begin{pmatrix}
s - 2k\pi & 0 \\
0 &  - s + 2k\pi
\end{pmatrix}  \, , 
\end{equation}
with $k \in \mathbbm{Z}$ and where $s$ defines the principal logarithm, i.e. the particular result obtained by setting $k=0$. With ${\cal S}^{(1)}$ at hand, one accesses ${\cal X}^{(1)}(\theta)$ over the entire auxiliary manifold via
\begin{equation}
{\cal X}^{(1)}(\theta) = e^{-i\theta {\cal S}^{(1)}} = 
\begin{pmatrix}
e^{-i\theta (s - 2k\pi)} & 0 \\
0 &  e^{+i\theta (s - 2k\pi)}
\end{pmatrix}  \, . \label{1stextraction}
\end{equation}
For $\theta=1$, Eq.~\ref{1stextraction} is consistent with the starting point given that ${\cal X}^{(1)}(1)={\cal X}(1)={\cal X}$. 

Let us now extract ${\cal S}$ again from ${\cal X}^{(1)}(\bar{\theta})$ with $\bar{\theta} \in ]0,1[$. Given that 
\begin{equation}
-(1+2k)\bar{\theta}\pi<\bar{\theta} (s - 2k\pi)\leq (1-2k)\bar{\theta}\pi \, ,
\end{equation}
there exists\footnote{The integer $\bar{k}$ obviously depends on $k$ and $\bar{\theta}$.} $\bar{k} \in \mathbbm{Z}$ such that
\begin{equation}
\bar{\theta} \bar{s} \equiv  \bar{\theta} (s - 2k\pi) + 2\bar{k}\pi \in ]-\pi,\pi] \, .
\end{equation}
With this definition of $\bar{s}$ and authorizing again the use of a non-principal logarithm, one has
\begin{align}
{\cal S}^{(\bar{\theta})} &= \frac{i}{\bar{\theta}} \log {\cal X}(\bar{\theta}) \nonumber \\
&= 
\begin{pmatrix}
\bar{s} - 2\frac{k'}{\bar{\theta}}\pi & 0 \\
0 &  -\bar{s} + 2\frac{k'}{\bar{\theta}}\pi
\end{pmatrix} \nonumber \\
&= 
\begin{pmatrix}
s - 2k\pi -2\frac{k'-\bar{k}}{\bar{\theta}}\pi & 0 \\
0 &  -s + 2k\pi +2\frac{k'-\bar{k}}{\bar{\theta}}\pi
\end{pmatrix}  \, , 
\end{align}
where $k' \in \mathbbm{Z}$. This new form of ${\cal S}$ leads in particular to
\begin{equation}
{\cal X}^{(\bar{\theta})}(1) =
\begin{pmatrix}
e^{-is}e^{+2i\frac{k'-\bar{k}}{\bar{\theta}}\pi} & 0 \\
0 &  e^{+is}e^{-2i\frac{k'-\bar{k}}{\bar{\theta}}\pi}
\end{pmatrix}  \, , 
\end{equation}
which is {\it not} consistent with the starting point, i.e. ${\cal X}^{(\bar{\theta})}(1)\neq {\cal X}$, as $(k'-\bar{k})/\bar{\theta}$ is not an integer in general. 

The above derivation proves that, although the solution to  Eq.~\ref{extraction1st} is a priori multivalued, demanding to have a consistent family of Bogoliubov transformations ${\cal X}(\theta)$ along the auxiliary manifold requires the sole use of the {\it principle} logarithm all throughout. Indeed, only at this condition one has an internally consistent approach manifested by the necessity to set $k=\bar{k}=k'=0$ to generate a family of transformations ${\cal X}(\theta)$ that is independent of the point $\bar{\theta} \in ]0,1]$ at which the intermediate matrix ${\cal S}$ is extracted.

\section{Symmetry transformation}
\label{symmetrytransformation}

\subsection{Introduction}

Unitary transformations of particular interest are transformations associated with symmetry groups that are subgroups of $U(N)$, the group of unitary transformations of a single-particle basis of dimension $N$. Proceeding to such a unitary transformation, states $| \Phi \rangle$ and $| \breve{\Phi} \rangle$ are related via Eq.~\ref{defphibreve} with an operator $S$ characterized by $s^{00}=s^{20}=s^{02}=0$, i.e. taking the particular form
\begin{eqnarray}  
S &=& \sum_{pq} s^{11}_{pq} c^{\dagger}_p c_q \nonumber \\
&=& \frac12 \Tr\left(s^{11}\right) 
+ \frac12  
\left(\,c^{\dagger} \hspace{0.2cm} c \,\right)
\begin{pmatrix}
s^{11} & 0 \\
0 &  -s^{11 \ast}
\end{pmatrix} 
\begin{pmatrix}
c \\
c^{\dagger}
\end{pmatrix}  \, . 
\end{eqnarray} 
In this case, the matrices defining $S$ in the quasiparticle basis of $| \Phi \rangle$ reduce to 
\begin{subequations}
\label{transfomatricesSsym}
\begin{align}
S^{00} &= \text{Tr}(s^{11} V^{\ast}V^T) \, , 
\label{transfomatricesSsym1} \\
S^{11} &= U^\dagger s^{11}U - V^\dagger s^{11 *} V \, , \label{transfomatricesSsym2} \\
S^{20} &= U^\dagger s^{11}V^* - V^\dagger s^{11 *}U^* \, , \label{transfomatricesSsym3} \\
S^{02} &= U^T s^{11 *} V - V^T s^{11} U   \, , \label{transfomatricesSsym4}
\end{align}
\end{subequations}

\subsection{Bogoliubov transformations within $\mathcal{M}[| \Phi \rangle,S]$}

Given the known operator $S$, the Bogoliubov transformations associated with states making up the auxiliary manifold $\mathcal{M}[| \Phi \rangle,S]$ can be obtained directly. 

The elementary commutator
\begin{align}
 \Big[S , c^{\dagger}_m \Big] &=  \sum_{kl} s^{11}_{kl} c^{\dagger}_k c_l c^{\dagger}_m - c^{\dagger}_m S \nonumber \\
                       &=  \sum_{kl} s^{11} c^{\dagger}_k  (\delta_{lm} - c^{\dagger}_m c_l ) - c^{\dagger}_m S \nonumber \\
                       &=  \sum_{k} s^{11}_{km} c^{\dagger}_k \, ,  
\end{align}
allows one, on the basis of Baker-Campbell-Hausdorff's identity, to write
\begin{widetext}
\begin{equation}
\begin{split}
e^{i \theta S} c^{\dagger}_m e^{- i \theta S} &= c^{\dagger}_m + \Big[ i\theta S , c^{\dagger}_m \Big] + \frac{1}{2!} \Big[ i\theta S , \Big[ i \theta S , c^{\dagger}_m \Big] \Big] +
      \frac{1}{3!} \Big[ i\theta S , \Big[ i\theta S , \Big[ i \theta S , c^{\dagger}_m \Big] \Big] \Big] + \ldots  \\
  &= c^{\dagger}_m + i\theta  \sum_{k} s^{11}_{km} c^{\dagger}_k + \frac{(i\theta)^2}{2!} \sum_{kl} s^{11}_{km} s^{11}_{lk} c^{\dagger}_l + \frac{(i\theta)^3}{3!} \sum_{kln} s^{11}_{km} s^{11}_{lk} s^{11}_{nl} c^{\dagger}_n + \ldots \\
  &= c^{\dagger}_m + i\theta  \sum_{k} s^{11}_{km} c^{\dagger}_k + \frac{(i\theta)^2}{2!} \sum_{l} (s^{11})^2_{lm} c^{\dagger}_l + \frac{(i\theta)^3}{3!} \sum_{n} (s^{11})^3_{nm} c^{\dagger}_n + \ldots \\
  &= \sum_{k} \big(e^{i\theta s^{11}} \big)_{km} c^{\dagger}_k \, .
\end{split}
\end{equation}
\end{widetext}
Similarly
\begin{equation}
 e^{i \theta S} c_m e^{- i \theta S} = \sum_{k} \big(e^{i\theta s^{11}} \big)^*_{km} c_k \, .
\end{equation}
With these relations at hand, one can compute
\begin{equation}
\begin{split}
 \beta_\mu^{\theta +} &\equiv  e^{i \theta S} \beta^{\dagger}_\mu e^{- i \theta S} \\
  &= \sum_{\nu} U_{\nu \mu} e^{i \theta S} c^{\dagger}_\nu e^{- i \theta S} + V_{\nu \mu} e^{i \theta S} c_\nu e^{- i \theta S} \\
 &= \sum_{\nu \lambda} U_{\nu \mu} \big(e^{i\theta s^{11}} \big)_{\lambda \nu} c^{\dagger}_\lambda + V_{\nu \mu} \big(e^{i\theta s^{11}} \big)^*_{\lambda \nu} c_\lambda \\ 
 &\equiv \sum_{\lambda} U^\theta_{\lambda \mu} c^{\dagger}_\lambda + V^\theta_{\lambda \mu} c_\lambda \, ,
\end{split}
\end{equation}
which provides the Bogoliubov transformation relating $| \Phi(\theta) \rangle$ to $| 0 \rangle$ under the form
\begin{subequations}
\label{transfoBogosysm}
\begin{align}
 U^\theta &\equiv e^{+i\theta s^{11}} U  \, , \\
 V^\theta &\equiv e^{-i\theta s^{11 *}} V   \, ,
\end{align}
\end{subequations}
leading, for $\theta= 1$, to
\begin{equation}
\breve{{\cal W}} \equiv
\begin{pmatrix}
\breve{U} & \breve{V}^{\ast} \\
\breve{V} &  \breve{U}^{\ast}
\end{pmatrix} 
=
\begin{pmatrix}
e^{+is^{11}} & 0 \\
0  &  e^{-i s^{11 *}}
\end{pmatrix} 
\begin{pmatrix}
U & V^{\ast} \\
V &  U^{\ast}
\end{pmatrix} \, .
\end{equation}
Thouless' matrix $Z^\theta$ is further obtained as
\begin{equation}
\begin{split}
 Z^\theta &\equiv V^{\theta *} \Big[ U^{\theta *} \Big]^{-1} \\ 
          &= e^{i\theta s^{11}} Z e^{i\theta s^{11 *}}\, .
\end{split}
\end{equation}
Next, the Bogoliubov transformation ${\cal X}(\theta)$ linking $\ket{\Phi}$ and $\ket{\Phi(\theta)}$ is obtained explicitly in terms of $s^{11}$ and $\theta$ from
\begin{subequations}
\begin{align}
\beta_{k_1} &= \sum_{k_2} A^{*}_{k_2k_1}(\theta) \, \beta^{\theta}_{k_2} 
 + B^{*}_{k_2k_1}(\theta) \,  \beta^{\theta \dagger}_{k_2} \, , \\
\beta_{k_1}^{\dagger} &= \sum_{k_2} A_{k_2k_1}(\theta) \, \beta^{\theta \dagger}_{k_2} 
 + B_{k_2k_1}(\theta) \,  \beta^{\theta}_{k_2} \, ,
\end{align}
\end{subequations}
with
\begin{subequations}
\begin{align}
A(\theta) &\equiv U^{\theta \dagger}U +  V^{\theta \dagger}V  \, , \\
B(\theta) &\equiv V^{\theta T}U +  U^{\theta T}V \, ,
\end{align}
\end{subequations}
which is a usable alternative to the exponential representation of Eq.~\ref{exprepresentationWtheta}. Eventually, the elementary contraction of interest is given by
\begin{align}
R^{--}[\langle \Phi |,| \Phi(\theta) \rangle] &= - B^\dagger (\theta) \big[ A^T(\theta) \big]^{-1} \\   
              &= -  V^\dagger \big[  1 -  Z^{-1} Z^{\theta} \big] \big[ 1 -  Z^* Z^{\theta} \big]^{-1}  ( U^{T} )^{-1} \, . \nonumber
\end{align}
With these quantities at hand, and remembering that $s^{02}=0$ in the present case, the norm kernel is obtained from Eq.~\ref{overlapreformulated} under the simplified form
\begin{align}
\frac{\langle \Phi | \breve{\Phi} \rangle}{\langle \Phi | \Phi \rangle} &=  e^{i S^{00}} e^{\frac{i}{2} \int_0^1 d\theta  \Tr \left(S^{02} R^{--}[\langle \Phi |,| \Phi(\theta) \rangle]\right)} \, . \label{overlapsym}
\end{align}

\section{Global gauge rotation}
\label{gaugerotation}

\subsection{Set up}

The particular case of global gauge rotation is obtained by employing $S \equiv \varphi A$ in the set of equations derived in App.~\ref{symmetrytransformation}, where $\varphi \in [0, 2\pi]$ denotes the gauge angle. 

One is interested in computing the overlap between the Bogoliubov state $| \Phi \rangle$ and its rotated partner $| \breve{\Phi} \rangle = e^{iA\varphi}| \Phi \rangle$.  The auxiliary manifold linking $| \Phi \rangle$ and $| \breve{\Phi} \rangle$ is\footnote{It is to be noted that $| \breve{\Phi} \rangle$, $\ket{\Phi(\theta)}$ and most quantities introduced below depend implicitly on the gauge angle $\varphi$, which is not to be confused with the angle $\theta$ running over the auxiliary manifold defined for any given value of $\varphi$.}
\begin{equation}
\mathcal{M}[| \Phi \rangle,S] \equiv \big\{ \ket{\Phi(\theta)} \equiv e^{i \theta \varphi A}  \ket{\Phi} \, , \, \theta \in [0, 1] \big\} \, .
\end{equation}

Inserting $s^{11}_{km} = \varphi \, \delta_{km}$ into Eq.~\ref{transfomatricesSsym}, the matrices defining $S$ in the quasiparticle basis of $| \Phi \rangle$ reduce to 
\begin{subequations}
\label{transfomatricesSsymgauge}
\begin{align}
S^{00} &\equiv \varphi A^{00} = \varphi \, \text{Tr}(V^{\ast}V^T) \, , \label{transfomatricesSsymgauge1} \\
S^{11} &\equiv \varphi A^{11} = \varphi \left[U^\dagger U - V^\dagger V\right] \, , \label{transfomatricesSsymgauge2} \\
S^{20} &\equiv \varphi A^{20} = \varphi \left[U^\dagger V^* - V^\dagger U^*\right] \, , \label{transfomatricesSsymgauge3} \\
S^{02} &\equiv \varphi A^{02} = \varphi \left[U^T  V - V^T  U \right]  \, . \label{transfomatricesSsymgauge4}
\end{align}
\end{subequations}
One obtains from Eq.~\ref{transfoBogosysm}
\begin{subequations}
\begin{align}
 U^\theta &\equiv e^{+i\theta\varphi} U  \, , \\
 V^\theta &\equiv e^{-i\theta\varphi} V   \, ,
\end{align}
\end{subequations}
leading, for $\theta=1$, to
\begin{equation}
\breve{{\cal W}} \equiv
\begin{pmatrix}
\breve{U} & \breve{V}^{\ast} \\
\breve{V} &  \breve{U}^{\ast}
\end{pmatrix} 
=
\begin{pmatrix}
e^{+i\varphi}U & e^{+i\varphi}V^{\ast} \\
e^{-i\varphi}V &  e^{-i\varphi}U^{\ast}
\end{pmatrix} \, . \label{gaugeBogobreve}
\end{equation}
Thouless' matrix $Z^\theta$ is further obtained as
\begin{equation}
\begin{split}
 Z^\theta &\equiv V^{\theta *} \Big[ U^{\theta *} \Big]^{-1} \\ 
          &= e^{2i\theta\varphi}Z  \, .
\end{split}
\end{equation}
The Bogoliubov transformation ${\cal X}(\theta)$ linking $\ket{\Phi}$ and $\ket{\Phi(\theta)}$ is built from
\begin{subequations}
\begin{align}
A(\theta) &\equiv e^{-i\theta\varphi} U^{\dagger}U +  e^{+i\theta\varphi}V^{\dagger}V  \, , \\
B(\theta) &\equiv e^{-i\theta\varphi}V^{T}U +  e^{+i\theta\varphi}U^{T}V \, ,
\end{align}
\end{subequations}
trivially providing ${\cal X}$ for $\theta=1$. The elementary contraction of interest takes the form
\begin{align}
 R^{--}[\langle \Phi |,| \Phi(\theta) \rangle] &= -  V^\dagger \big( 1 - e^{2i\theta\varphi} \big) \nonumber \\
 & \,\,\,\,\,\, \times \big( \mathds{1} - e^{2i\theta\varphi} Z^* Z \big)^{-1} ( U^T )^{-1} \, . 
\end{align}
Eventually, the norm kernel is
\begin{align}
\frac{\langle \Phi | \breve{\Phi} \rangle}{\langle \Phi | \Phi \rangle} &=  e^{i \varphi \, A^{00}} e^{\frac{i}{2} \int_0^1 d\theta  \Tr \left(\varphi A^{02} R^{--}[\langle \Phi |,| \Phi(\theta) \rangle]\right)} \, , \label{overlapsymgauge}
\end{align}
and rewrites under the change of variable $\phi\equiv \varphi \, \theta$ as
\begin{subequations}
\label{overlapsymgauge2}
\begin{align}
\frac{\langle \Phi | \breve{\Phi} \rangle}{\langle \Phi | \Phi \rangle} &= e^{i  \int_0^{\varphi} d\phi  \frac{\langle \Phi |A| \Phi(\phi/\varphi) \rangle}{\langle \Phi |\Phi(\phi/\varphi) \rangle}} \label{overlapsymgauge2A} \\
&= e^{i \varphi \, A^{00}} e^{\frac{i}{2} \int_0^{\varphi} d\phi  \Tr \left(A^{02} R^{--}(\phi)\right)} \, , \label{overlapsymgauge2B}
\end{align}
\end{subequations}
where
\begin{align}
 R^{--}(\phi) &\equiv R^{--}[\langle \Phi |,| \Phi(\phi/\varphi) \rangle] \label{eq:rmmsecond2}  \\
 &=-  V^\dagger \big( 1 - e^{2i\phi} \big) \big( \mathds{1} - e^{2i\phi} Z^* Z \big)^{-1} ( U^T )^{-1} \, . \nonumber
\end{align}
Equation~\ref{overlapsymgauge2B} is the formula of the norm kernel given just below Eq.~126b of Ref.~\cite{Duguet:2015yle}.

\subsection{Canonical basis}

In order to analytically scrutinize expression~\ref{overlapsymgauge2}, we work from there on in the canonical basis. Accordingly, the Bogoliubov transformation characterizing $| \Phi \rangle$ is considered to be given under the simple block diagonal form with real $2\times 2$ blocks defined through\footnote{We presently deal with even-number parity states for simplicity. The discussion can however be easily extended to Bogoliubov states obtained via an arbitrary even or odd number of quasi-particle excitations on top of such vacua.}
\begin{subequations}
\label{gaugeBogocano}
\begin{alignat}{2}
 V_{k_1 k_2}   &= +v_{k_1} \delta_{k_2 \bar{k}_1} &&= V_{k_1 \bar{k}_1} \, , \\
 V_{k_1 k_2}^T &= -v_{k_1} \delta_{k_2 \bar{k}_1} &&= V_{\bar{k}_1 k_1} \, , \\
 U_{k_1 k_2}   &= +u_{k_1} \delta_{k_2      k _1} &&= U_{\bar{k}_1 \bar{k}_1}  \, , \\
 U_{k_1 k_2}^T &= +u_{k_1} \delta_{k_2      k _1} &&= U^T_{\bar{k}_1 \bar{k}_1}  \, , 
\end{alignat}
\end{subequations}
where $\bar{k}_1$ represents the conjugate partner of $k_1$ and where Eq.~\ref{unitarityA} reduces to
\begin{equation}
  u_{k_1}^2 + v_{k_1}^2 = 1 \, .
\end{equation}
The canonical representation of the Bogoliubov transformation associated with state $| \breve{\Phi} \rangle$ is easily deduced from Eq.~\ref{gaugeBogobreve}.

In the canonical quasi-particle basis of $| \Phi \rangle$, the non-zero off-diagonal contraction of interest takes the simplified form
\begin{equation}
 R^{--}_{k_1 k_2}(\phi) = - \frac{u_{k_1} v_{k_1}\big( e^{2i\phi} - 1 \big)}{u_{k_1}^2 + v_{k_1}^2 e^{2i\phi}}  \delta_{k_2 \bar{k}_1}  \, ,  \label{eq:rmmfin} 
\end{equation}
and vanishes for $\phi=k\pi$, with $k \in \mathbbm{Z}$. One further observes that $R^{--}_{k_1\bar{k}_1}(\phi)$ diverges for $\phi = \pi/2 + k\pi$ whenever $u_{k_1}^2 = v_{k_1}^2 = 0.5$~\cite{tajima92a,donau98,almehed01a,anguiano01b,dobaczewski07,Bender:2008rn}.

In the canonical quasi-particle basis of $| \Phi \rangle$, Eq.~\ref{transfomatricesSsymgauge} is computed thanks to
\begin{subequations}
\label{transfomatricesSsymgaugecano}
\begin{align}
A^{00}  &=  2 \sum_{k_1 > 0} v_{k_1}^2 \, , \label{transfomatricesSsymgaugecano1} \\
A^{11}_{k_1 k_2} &= \big( u_{k_1}^2 - v_{k_1}^2 \big) \delta_{k_1 k_2} \, , \label{transfomatricesSsymgaugecano2} \\
A^{20}_{k_1 k_2} &=  2 u_{k_1} v_{k_1} \delta_{k_2 \bar{k}_1} \, , \label{transfomatricesSsymgaugecano3} \\
A^{02}_{k_1 k_2} &=  2 u_{k_1} v_{k_1} \delta_{k_2 \bar{k}_1} \, , \label{transfomatricesSsymgaugecano4}
\end{align}
\end{subequations}
where we have used that $v_{k_1}^2 = v_{\bar{k}_1}^2$ to reduce the sum to only half of the basis of paired particles, denoted by the label $\sum_{k_1 > 0}$.

\subsection{Norm kernel}

\subsubsection{Reference formula}
\label{sec:anafor}

It is possible to explicitly represent the vacua associated with the canonical Bogoliubov transformations introduced above under the form
\begin{subequations}
\begin{align}
| \Phi \rangle &\equiv \prod_{k_1>0} (u_{k_1} + v_{k_1} c^{\dagger}_{k_1} c^{\dagger}_{\bar{k}_1}) | 0 \rangle \,\, , \\
| \breve{\Phi} \rangle &= \prod_{k_1>0}(u_{k_1} + e^{2i\varphi} v_{k_1} c^{\dagger}_{k_1} c^{\dagger}_{\bar{k}_1}) | 0 \rangle \,\, .
\end{align}
\end{subequations}
This representation is consistent with the phase convention $\text{Arg} (\langle 0 | \Phi \rangle) = \text{Arg} (\langle 0 | \breve{\Phi} \rangle) = 0$. The two states are explicitly normalized and their overlap is
\begin{align}
 \label{eq:overcano}
\frac{\langle \Phi | \breve{\Phi} \rangle}{\langle \Phi | \Phi \rangle}  &= \langle 0 | \prod_{k_1>0} (u_{k_1} + v_{k_1}  c_{\bar{k}_1} c_{k_1})(u_{k_1} + e^{2i\varphi} v_{k_1} c^{\dagger}_{k_1} c^{\dagger}_{\bar{k}_1}) | 0 \rangle \nonumber \\
&= \prod_{k_1 > 0} \big( u_{k_1}^2 + v_{k_1}^2 e^{2i\varphi} \big) \\
&\equiv \prod_{k_1 > 0} z_{k_1}(\varphi) \, , \nonumber
\end{align}
where the result is obtained via the application of elementary anticommutation rules. 
The polar form of the overlap is trivially obtained by setting $z_{k_1}(\varphi)\equiv r_{k_1}(\varphi)  e^{i \theta_{k_1}(\varphi)}$ where
\begin{subequations}
\begin{align}
r_{k_1}(\varphi) &\equiv \sqrt{u_{k_1}^4 + v_{k_1}^4 + 2 u_{k_1}^2 v_{k_1}^2 \cos(2\varphi)} \, , \\
\theta_{k_1}(\varphi) &\equiv \arctan \Big( \frac{ v_{k_1}^2 \sin(2\varphi)}{u_{k_1}^2 + v_{k_1}^2 \cos(2\varphi)} \Big) \, .
\end{align}
\end{subequations}

Formula~\ref{eq:overcano} testifies that the norm overlap is strictly 0 at $\varphi = \pi/2$ as soon as a conjugate pair is such that $u_{k_1}^2=v_{k_1}^2=0.5$, i.e. the norm overlap is zero due to the fact that $z_{k_1}(\pi/2)=0$ in this case. As a matter of fact, the contribution of such a pair to the norm kernel reads, as a function of $\varphi$, as
\begin{align}
z_{k_1}(\varphi) 
&= \frac{1}{2} \big( 1 + e^{2i\varphi} \big) = e^{i\varphi} \cos (\varphi) \, . \label{symmetry1}
\end{align}

\subsubsection{Integral formula}
\label{sec:deriva0}

Employing the exponential formula consistent with the phase convention $\text{Arg} (\langle 0 | \Phi \rangle) = \text{Arg} (\langle 0 | \breve{\Phi} \rangle) = 0$, the norm overlap is obtained from Eqs.~\ref{overlapsymgauge2},~\ref{eq:rmmfin} and~\ref{transfomatricesSsymgaugecano} as
\begin{equation}
\label{analytic1}
\begin{split}
\frac{\langle \Phi | \breve{\Phi} \rangle}{\langle \Phi | \Phi \rangle}  
   &= e^{\sum_{k_1 > 0} \int_0^\varphi d\phi \, \frac{2 i v_{k_1}^2 e^{2i\phi}}{u_{k_1}^2 + v_{k_1}^2 e^{2i\phi}}} \\
   &= e^{\sum_{k_1 > 0} \ln\big(u_{k_1}^2 + v_{k_1}^2 e^{2i\varphi} \big)} \\
   &= \prod_{k_1 > 0} \big( u_{k_1}^2 + v_{k_1}^2 e^{2i\varphi} \big)  \, ,
\end{split}
\end{equation}
which matches the result of the direct calculation given in Eq.~\ref{eq:overcano}. 

In fact, the {\it derivation} in Eq.~\ref{analytic1} is only valid under the assumption that no factor $z_{k_1}(\phi)$ is zero over the integration interval $\phi \in [0,\varphi]$, i.e. under the assumption that the overlap between the initial state $| \Phi \rangle$ and any intermediate gauge-rotated states is different from zero over the interval $\phi \in [0,\varphi]$. Whenever (at least) one conjugate pair is characterized by $u_{k_1}^2=v_{k_1}^2=0.5$, the fact that $z_{k_1}(\pi/2)=0$ limits the validity of Eq.~\ref{analytic1} over the limited interval $\varphi \in [0,\pi/2[$. Let us now discuss what happens specifically for $\varphi = \pi/2$ and $\varphi > \pi/2$ in this situation.

\subsubsection{Orthogonality}
\label{sec:divergence}

Given that Eq.~\ref{eq:overcano} is properly recovered from Eq.~\ref{overlapsymgauge2} for $\varphi \in [0,\pi/2[$ whenever there exists $k_1$ such that $z_{k_1}(\pi/2)=0$, one can test whether the correct value is obtained in the limit $\varphi \rightarrow  \pi/2$. The zero of the norm overlap being induced by the sole pair $(k_1,\bar{k}_1)$, it is sufficient to focus on its contribution.

We restart from Eq.~\ref{overlapsymgauge2B} and separate the real and imaginary parts of $R^{--}_{\bar{k}_1 k_1} (\phi)$
\begin{equation}
\begin{split}
 \label{eq:rmmcossin}
 R^{--}_{\bar{k}_1 k_1} (\phi) &= \frac{u_{k_1} v_{k_1}\big( e^{2i\phi} - 1 \big)}{u_{k_1}^2 + v_{k_1}^2 e^{2i\phi}} \\ 
                               &=\frac{u_{k_1} v_{k_1} \Big[ \big(u_{k_1}^2 - v_{k_1}^2\big) (\cos(2\phi) - 1) + i \sin(2\phi) \Big]}{u_{k_1}^4 + v_{k_1}^4 + 2 u_{k_1}^2 v_{k_1}^2 \cos(2\phi)} \, .
\end{split}
\end{equation}
Setting $u_{k_1}^2=v_{k_1}^2=0.5$, the contribution of the pair of interest to the norm kernel reads as
\begin{align}
\left(\frac{\langle \Phi | \breve{\Phi} \rangle}{\langle \Phi | \Phi \rangle}\right)_{k_1\bar{k}_1} 
&= e^{i\varphi} e^{- \int_0^\varphi d\phi \frac{\sin(2\phi) }{1 + \cos(2\phi)}} \nonumber \\
&= e^{i\varphi} e^{- \int_0^\varphi d\phi \tan(\phi)} \label{formulaortho} \\
&= e^{i\varphi} e^{\ln\big(|\cos(\varphi)|\big)}  \nonumber  \\
&= e^{i\varphi} |\cos(\varphi)|  \, .  \nonumber 
\end{align}
Taking the limit for $\varphi \rightarrow  \pi/2$ smoothly leads to the expected cancellation of the norm overlap. This proves that, in spite of the derivation of the exponential formula being only valid over the restricted interval $\varphi \in [0,\pi/2[$, it safely provides the nullity of the overlap when $\varphi \rightarrow  \pi/2$. This correct limit is obtained from the divergence of the imaginary part of the elementary contraction $R^{--}_{\bar{k}_1 k_1} (\phi)$ whose sign is such that  the exponential properly drives the overlap to zero.

\subsubsection{Going through zeros of the overlap}
\label{throughzero}

While the cancellation of the overlap is smoothly obtained for $\varphi =  \pi/2$, one may wonder if the analytical form obtained in Eq.~\ref{formulaortho} can be safely extrapolated to $\varphi >  \pi/2$, in spite of the fact that the derivation is not valid in this case. As a matter of fact, Eq.~\ref{formulaortho} leads to a sign difference for $\varphi > \pi/2$ as compared to the correct formula, as can be understood from Eq.~\ref{symmetry1}. More specifically, the extrapolation of the exponential formula to the interval $[\pi/2,3\pi/2]$ differs from the correct value by a sign $(-1)^{p}$, where $p$ denotes the number of conjugated pairs characterized by $u_k=v_k=0.5$. Eventually, the sign is correct again when going through the next zero of the overlap, i.e. on the interval $[3\pi/2,2\pi]$, independently of $p$ given that $|\cos(\varphi)|=\cos(\varphi)$ on such an interval, just as it is the case over the interval $[0,\pi/2]$.

\section{Particle-number restoration}
\label{appendixPNR} 

To support the discussion provided in Sec.~\ref{symmetryrestoration} regarding the computation of the norm matrix at play in symmetry restoration calculations, we presently detail the restoration of good particle number associated with global gauge symmetry. In particular, we illustrate that the symmetry restoration can be equally achieved via the use of a projection operator or via a diagonalization method, both methods making a different use of the associated norm matrix. 

\subsection{Particle-number restoration as a projection}

Given the particle-number breaking state $| \Phi \rangle$, an eigenstate of $A$ with eigenvalue $\text{A}$ is obtained as
\begin{align}
| \Phi^{\text{A}} \rangle &\equiv P^{\text{A}} | \Phi \rangle \equiv \int_{0}^{2\pi} \frac{d\varphi}{2\pi} \, e^{-i \varphi \text{A}} \, | \Phi(\varphi) \rangle \, , \label{projection}
\end{align}
as demonstrated by
\begin{align}
A | \Phi^{\text{A}} \rangle &= -i \int_{0}^{2\pi} \frac{d\varphi}{2\pi} \, e^{-i \varphi \text{A}} \, \frac{d}{d\varphi} | \Phi(\varphi) \rangle \nonumber \\
& = + i \int_{0}^{2\pi} \frac{d\varphi}{2\pi} \, \frac{d}{d\varphi}\Big(e^{-i \varphi \text{A}}\Big)  | \Phi(\varphi) \rangle  \nonumber \\
&= \text{A} | \Phi^{\text{A}} \rangle \, .
\end{align}
The particle-number restored energy is computed as
\begin{subequations}
\label{projEannexe}
\begin{align}
\text{E}^{\text{A}} &\equiv \frac{\langle \Phi^{\text{A}}  | H | \Phi^{\text{A}}  \rangle}{\langle \Phi^{\text{A}}  | \Phi^{\text{A}}  \rangle} \\
&= \frac{\int_{0}^{2\pi} \!d\varphi \, e^{-i \varphi \text{A}} \, \langle \Phi | H | \Phi(\varphi) \rangle}{\int_{0}^{2\pi} \!d\varphi \, e^{-i \varphi \text{A}} \, \langle \Phi  | \Phi(\varphi) \rangle}    \, ,
\end{align}
\end{subequations}
where the fact that $P^{\text{A}}$ is a hermitian projector ($(P^{\text{A}})^\dagger=P^{\text{A}}$ and $(P^{\text{A}})^2=P^{\text{A}}$) commuting with the Hamiltonian ($[H,P^{\text{A}}]=0$) was used. Interestingly, the calculation of particle-number restored observables associated with any hermitian operator $O$ ($O^\dagger = O$) of rank\footnote{An operator of rank $\text{r}$ with respect to the $U(1)$ group  is such that 
\begin{equation}
\, P^{\text{A}} O = O  P^{\text{A} - \text{r}} \, ,  \,\, \forall \, \text{A} \in \mathbb{N}  \, .
\end{equation}
}
$r$ only makes use of the first row of the norm matrix.

\subsection{Particle-number restoration as a diagonalization}

Alternatively, the symmetry restoration can be achieved via a diagonalization of the nuclear Hamiltonian within the subspace spanned
by the MR set of gauge rotated states, namely the linear span  
\begin{equation}
\text{span}(\mathcal{M}_{U(1)}) \equiv \Bigg\{ \int_{0}^{2\pi} \frac{d\varphi}{2\pi} \, f(\varphi) | \Phi(\varphi) \rangle \, , \, f(\varphi) \in L^2([0,2\pi]) \Bigg\}  \nonumber
\end{equation}
where $L^2([0,2\pi])$ is the space of square-integrable functions over $[0,2\pi]$. 

To do so, the un-normalized projected states $| \Phi^{\text{A}} \rangle$ defined in Eq.~\ref{projection} can be first recovered via the diagonalization of the norm matrix. Expanding the Bogoliubov state $| \Phi \rangle$ over normalized eigenstates of $A$
\begin{equation}
\label{expandphi}
| \Phi \rangle \equiv  \sum_{\text{A}} c_{\text{A}} \, | \Lambda^{\text{A}} \rangle \, ,
\end{equation}
the norm kernel is rewritten as
\begin{equation}
\label{expandnormmatrix}
\langle \Phi(\varphi) |  \Phi(\varphi') \rangle = \sum_{\text{A}} |c_{\text{A}}|^{2} \, e^{i (\varphi'-\varphi) \text{A}}  \, .
\end{equation}
The goal is to find the weights $f_k(\varphi)$ of the states diagonalizing the norm matrix, i.e. 
\begin{equation}
\label{diagonormmatrix}
\int_{0}^{2\pi} \frac{d\varphi'}{2\pi} \, \langle \Phi(\varphi) |  \Phi(\varphi') \rangle  \, f_{k}(\varphi') =  n_k  \, f_{k}(\varphi) \, ,
\end{equation}
where $n_k$ denotes the corresponding eigenvalues. Employing Eq.~ \ref{expandnormmatrix} and Fourier expanding the weights according to
\begin{equation}
\label{expandweight}
f_{k}(\varphi) \equiv  \sum_{\text{A}}  f_k^{-\text{A}}  \, e^{i \varphi \text{A}} \, ,
\end{equation}
Eq.~\ref{diagonormmatrix} can be easily shown to be equivalent to the set of equations
\begin{equation}
\label{diagonormmatrixrewritten}
f_k^{\text{A}}(|c_{\text{A}}|^{2}-n_k) =0 \,\, , \,\, \forall \, \text{A}  \in \mathbb{N} \, .
\end{equation}
The solutions of Eq.~\ref{diagonormmatrixrewritten} are
\begin{subequations}
\label{solutions}
\begin{align}
f_k^{\text{A}} &= \delta_{\text{A}k} \, , \label{solutions1} \\
f_{\text{A}}(\varphi) &= e^{-i \varphi \text{A}} \, , \label{solutions2} \\
n_{\text{A}} &= |c_{\text{A}}|^{2}   \, ,  \label{solutions3} 
\end{align}
\end{subequations}
which are nothing but the states $| \Phi^{\text{A}} \rangle$ introduced in Eq.~\ref{projection}. The corresponding normalized eigenstates of $A$ read as
\begin{equation}
\label{projectedstates}
| \Lambda^{\text{A}} \rangle \equiv \frac{1}{\sqrt{n_{\text{A}}}} | \Phi^{\text{A}} \rangle \, .
\end{equation}
Equation~\ref{solutions3} in particular demonstrates that the eigenvalues of the norm overlap matrix are nothing but the probability to find the normalized eigenstates $| \Lambda^{\text{A}} \rangle$ in the Bogoliubov state $|  \Phi \rangle$
\begin{equation}
\label{componenteigenstates}
n_{\text{A}} = |c_{\text{A}}|^{2} = \langle \Phi | P^{\text{A}} |  \Phi \rangle \, .
\end{equation}
Given that
\begin{equation}
\label{energymatrix}
\langle \Phi(\varphi) | H | \Phi(\varphi') \rangle = \sum_{\text{A}} \text{E}_{\text{A}} \, |c_{\text{A}}|^{2} \, e^{i (\varphi'-\varphi) \text{A}} \, ,
\end{equation}
where $\text{E}_{\text{A}} \equiv \langle \Lambda^{\text{A}} | H | \Lambda^{\text{A}} \rangle$, the diagonalization of the Hamiltonian matrix ${\cal H}_{\mathcal{M}_{U(1)}}$ provides the same eigenstates as the norm matrix and extracts the symmetry-restored energies $\text{E}_{\text{A}}$ as eigenvalues (up to the factor $|c_{\text{A}}|^{2}$ which is extracted first via the diagonalization of the norm matrix). Interestingly, the calculation of particle-number restored observables makes use of the complete norm and operator matrices. This is to be contrasted with the method based on the projector that only makes use of their first row and is thus more economical in practical calculations.

\end{appendix}

\bibliography{overlap}

\begin{thebibliography}{23}%
\makeatletter
\providecommand \@ifxundefined [1]{%
 \@ifx{#1\undefined}
}%
\providecommand \@ifnum [1]{%
 \ifnum #1\expandafter \@firstoftwo
 \else \expandafter \@secondoftwo
 \fi
}%
\providecommand \@ifx [1]{%
 \ifx #1\expandafter \@firstoftwo
 \else \expandafter \@secondoftwo
 \fi
}%
\providecommand \natexlab [1]{#1}%
\providecommand \enquote  [1]{``#1''}%
\providecommand \bibnamefont  [1]{#1}%
\providecommand \bibfnamefont [1]{#1}%
\providecommand \citenamefont [1]{#1}%
\providecommand \href@noop [0]{\@secondoftwo}%
\providecommand \href [0]{\begingroup \@sanitize@url \@href}%
\providecommand \@href[1]{\@@startlink{#1}\@@href}%
\providecommand \@@href[1]{\endgroup#1\@@endlink}%
\providecommand \@sanitize@url [0]{\catcode `\\12\catcode `\$12\catcode
  `\&12\catcode `\#12\catcode `\^12\catcode `\_12\catcode `\%12\relax}%
\providecommand \@@startlink[1]{}%
\providecommand \@@endlink[0]{}%
\providecommand \url  [0]{\begingroup\@sanitize@url \@url }%
\providecommand \@url [1]{\endgroup\@href {#1}{\urlprefix }}%
\providecommand \urlprefix  [0]{URL }%
\providecommand \Eprint [0]{\href }%
\providecommand \doibase [0]{http://dx.doi.org/}%
\providecommand \selectlanguage [0]{\@gobble}%
\providecommand \bibinfo  [0]{\@secondoftwo}%
\providecommand \bibfield  [0]{\@secondoftwo}%
\providecommand \translation [1]{[#1]}%
\providecommand \BibitemOpen [0]{}%
\providecommand \bibitemStop [0]{}%
\providecommand \bibitemNoStop [0]{.\EOS\space}%
\providecommand \EOS [0]{\spacefactor3000\relax}%
\providecommand \BibitemShut  [1]{\csname bibitem#1\endcsname}%
\let\auto@bib@innerbib\@empty
\bibitem [{\citenamefont {Ring}\ and\ \citenamefont {Schuck}(1980)}]{ring80a}%
  \BibitemOpen
  \bibfield  {author} {\bibinfo {author} {\bibfnamefont {P.}~\bibnamefont
  {Ring}}\ and\ \bibinfo {author} {\bibfnamefont {P.}~\bibnamefont {Schuck}},\
  }\href@noop {} {\emph {\bibinfo {title} {The Nuclear Many-Body Problem}}}\
  (\bibinfo  {publisher} {Springer-Verlag},\ \bibinfo {address} {New-York},\
  \bibinfo {year} {1980})\BibitemShut {NoStop}%
\bibitem [{\citenamefont {Bender}\ \emph {et~al.}(2003)\citenamefont {Bender},
  \citenamefont {Heenen},\ and\ \citenamefont {Reinhard}}]{bender03b}%
  \BibitemOpen
  \bibfield  {author} {\bibinfo {author} {\bibfnamefont {M.}~\bibnamefont
  {Bender}}, \bibinfo {author} {\bibfnamefont {P.-H.}\ \bibnamefont {Heenen}},
  \ and\ \bibinfo {author} {\bibfnamefont {P.-G.}\ \bibnamefont {Reinhard}},\
  }\href@noop {} {\bibfield  {journal} {\bibinfo  {journal} {Rev. Mod. Phys.}\
  }\textbf {\bibinfo {volume} {75}},\ \bibinfo {pages} {121} (\bibinfo {year}
  {2003})}\BibitemShut {NoStop}%
\bibitem [{\citenamefont {Duguet}(2014)}]{Duguet:2013dga}%
  \BibitemOpen
  \bibfield  {author} {\bibinfo {author} {\bibfnamefont {T.}~\bibnamefont
  {Duguet}},\ }\href@noop {} {\bibfield  {journal} {\bibinfo  {journal} {Lect.
  Notes Phys.}\ }\textbf {\bibinfo {volume} {879}},\ \bibinfo {pages} {293}
  (\bibinfo {year} {2014})}\BibitemShut {NoStop}%
\bibitem [{\citenamefont {Egido}(2016)}]{Egido:2016bdz}%
  \BibitemOpen
  \bibfield  {author} {\bibinfo {author} {\bibfnamefont {J.~L.}\ \bibnamefont
  {Egido}},\ }\href@noop {} {\bibfield  {journal} {\bibinfo  {journal} {Phys.
  Scripta}\ }\textbf {\bibinfo {volume} {91}},\ \bibinfo {pages} {073003}
  (\bibinfo {year} {2016})}\BibitemShut {NoStop}%
\bibitem [{\citenamefont {Blaizot}\ and\ \citenamefont
  {Ripka}(1986)}]{blaizot86}%
  \BibitemOpen
  \bibfield  {author} {\bibinfo {author} {\bibfnamefont {J.}~\bibnamefont
  {Blaizot}}\ and\ \bibinfo {author} {\bibfnamefont {G.}~\bibnamefont
  {Ripka}},\ }\href@noop {} {\emph {\bibinfo {title} {Quantum Theory of Finite
  Systems}}}\ (\bibinfo  {publisher} {MIT Press},\ \bibinfo {address}
  {Cambridge, Massachusetts},\ \bibinfo {year} {1986})\BibitemShut {NoStop}%
\bibitem [{\citenamefont {Robledo}(2009)}]{Robledo:2009yd}%
  \BibitemOpen
  \bibfield  {author} {\bibinfo {author} {\bibfnamefont {L.~M.}\ \bibnamefont
  {Robledo}},\ }\href@noop {} {\bibfield  {journal} {\bibinfo  {journal} {Phys.
  Rev.}\ }\textbf {\bibinfo {volume} {C79}},\ \bibinfo {pages} {021302}
  (\bibinfo {year} {2009})}\BibitemShut {NoStop}%
\bibitem [{\citenamefont {Robledo}(2011)}]{Robledo:2011ce}%
  \BibitemOpen
  \bibfield  {author} {\bibinfo {author} {\bibfnamefont {L.~M.}\ \bibnamefont
  {Robledo}},\ }\href@noop {} {\bibfield  {journal} {\bibinfo  {journal} {Phys.
  Rev.}\ }\textbf {\bibinfo {volume} {C84}},\ \bibinfo {pages} {014307}
  (\bibinfo {year} {2011})}\BibitemShut {NoStop}%
\bibitem [{\citenamefont {Avez}\ and\ \citenamefont
  {Bender}(2012)}]{Avez:2011wr}%
  \BibitemOpen
  \bibfield  {author} {\bibinfo {author} {\bibfnamefont {B.}~\bibnamefont
  {Avez}}\ and\ \bibinfo {author} {\bibfnamefont {M.}~\bibnamefont {Bender}},\
  }\href@noop {} {\bibfield  {journal} {\bibinfo  {journal} {Phys. Rev.}\
  }\textbf {\bibinfo {volume} {C85}},\ \bibinfo {pages} {034325} (\bibinfo
  {year} {2012})}\BibitemShut {NoStop}%
\bibitem [{\citenamefont {Duguet}\ and\ \citenamefont
  {Signoracci}(2017)}]{Duguet:2015yle}%
  \BibitemOpen
  \bibfield  {author} {\bibinfo {author} {\bibfnamefont {T.}~\bibnamefont
  {Duguet}}\ and\ \bibinfo {author} {\bibfnamefont {A.}~\bibnamefont
  {Signoracci}},\ }\href@noop {} {\bibfield  {journal} {\bibinfo  {journal} {J.
  Phys.}\ }\textbf {\bibinfo {volume} {G44}},\ \bibinfo {pages} {015103}
  (\bibinfo {year} {2017})}\BibitemShut {NoStop}%
\bibitem [{\citenamefont {Arthuis}\ \emph {et~al.}(2017)\citenamefont
  {Arthuis}, \citenamefont {Bally}, \citenamefont {Ebran},\ and\ \citenamefont
  {Duguet}}]{arthuis17a}%
  \BibitemOpen
  \bibfield  {author} {\bibinfo {author} {\bibfnamefont {P.}~\bibnamefont
  {Arthuis}}, \bibinfo {author} {\bibfnamefont {B.}~\bibnamefont {Bally}},
  \bibinfo {author} {\bibfnamefont {J.~P.}\ \bibnamefont {Ebran}}, \ and\
  \bibinfo {author} {\bibfnamefont {T.}~\bibnamefont {Duguet}},\ }\href@noop {}
  {} (\bibinfo {year} {2017}),\ \Eprint {http://arxiv.org/abs/unpublished}
  {unpublished} \BibitemShut {NoStop}%
\bibitem [{\citenamefont {Ring}\ and\ \citenamefont {Schuck}(1977)}]{ring77a}%
  \BibitemOpen
  \bibfield  {author} {\bibinfo {author} {\bibfnamefont {P.}~\bibnamefont
  {Ring}}\ and\ \bibinfo {author} {\bibfnamefont {P.}~\bibnamefont {Schuck}},\
  }\href@noop {} {\bibfield  {journal} {\bibinfo  {journal} {Nucl. Phys.}\
  }\textbf {\bibinfo {volume} {A292}},\ \bibinfo {pages} {20} (\bibinfo {year}
  {1977})}\BibitemShut {NoStop}%
\bibitem [{\citenamefont {Hara}\ and\ \citenamefont {Iwasaki}(1979)}]{hara79a}%
  \BibitemOpen
  \bibfield  {author} {\bibinfo {author} {\bibfnamefont {K.}~\bibnamefont
  {Hara}}\ and\ \bibinfo {author} {\bibfnamefont {S.}~\bibnamefont {Iwasaki}},\
  }\href@noop {} {\bibfield  {journal} {\bibinfo  {journal} {Nucl. Phys.}\
  }\textbf {\bibinfo {volume} {A332}},\ \bibinfo {pages} {61} (\bibinfo {year}
  {1979})}\BibitemShut {NoStop}%
\bibitem [{\citenamefont {Duguet}(2015)}]{Duguet:2014jja}%
  \BibitemOpen
  \bibfield  {author} {\bibinfo {author} {\bibfnamefont {T.}~\bibnamefont
  {Duguet}},\ }\href@noop {} {\bibfield  {journal} {\bibinfo  {journal} {J.
  Phys.}\ }\textbf {\bibinfo {volume} {G42}},\ \bibinfo {pages} {025107}
  (\bibinfo {year} {2015})}\BibitemShut {NoStop}%
\bibitem [{\citenamefont {Thouless}(1960)}]{thouless60}%
  \BibitemOpen
  \bibfield  {author} {\bibinfo {author} {\bibfnamefont {D.~J.}\ \bibnamefont
  {Thouless}},\ }\href@noop {} {\bibfield  {journal} {\bibinfo  {journal} {Ann.
  Phys.}\ }\textbf {\bibinfo {volume} {10}},\ \bibinfo {pages} {553} (\bibinfo
  {year} {1960})}\BibitemShut {NoStop}%
\bibitem [{\citenamefont {Balian}\ and\ \citenamefont
  {Br{\'e}zin}(1969)}]{balian69a}%
  \BibitemOpen
  \bibfield  {author} {\bibinfo {author} {\bibfnamefont {R.}~\bibnamefont
  {Balian}}\ and\ \bibinfo {author} {\bibfnamefont {E.}~\bibnamefont
  {Br{\'e}zin}},\ }\href@noop {} {\bibfield  {journal} {\bibinfo  {journal}
  {Nuovo Cimento}\ }\textbf {\bibinfo {volume} {64}},\ \bibinfo {pages} {37}
  (\bibinfo {year} {1969})}\BibitemShut {NoStop}%
\bibitem [{\citenamefont {Onishi}\ and\ \citenamefont
  {Yoshida}(1966)}]{onishi66}%
  \BibitemOpen
  \bibfield  {author} {\bibinfo {author} {\bibfnamefont {N.}~\bibnamefont
  {Onishi}}\ and\ \bibinfo {author} {\bibfnamefont {S.}~\bibnamefont
  {Yoshida}},\ }\href@noop {} {\bibfield  {journal} {\bibinfo  {journal} {Nucl.
  Phys.}\ }\textbf {\bibinfo {volume} {80}},\ \bibinfo {pages} {367} (\bibinfo
  {year} {1966})}\BibitemShut {NoStop}%
\bibitem [{\citenamefont {Wimmer}(2012)}]{wimmer12a}%
  \BibitemOpen
  \bibfield  {author} {\bibinfo {author} {\bibfnamefont {M.}~\bibnamefont
  {Wimmer}},\ }\href@noop {} {\bibfield  {journal} {\bibinfo  {journal} {ACM
  Trans. Math. Software}\ }\textbf {\bibinfo {volume} {38}},\ \bibinfo {pages}
  {10} (\bibinfo {year} {2012})}\BibitemShut {NoStop}%
\bibitem [{\citenamefont {Tajima}\ \emph {et~al.}(1992)\citenamefont {Tajima},
  \citenamefont {Flocard}, \citenamefont {Bonche}, \citenamefont
  {Dobaczewski},\ and\ \citenamefont {Heenen}}]{tajima92a}%
  \BibitemOpen
  \bibfield  {author} {\bibinfo {author} {\bibfnamefont {N.}~\bibnamefont
  {Tajima}}, \bibinfo {author} {\bibfnamefont {H.}~\bibnamefont {Flocard}},
  \bibinfo {author} {\bibfnamefont {P.}~\bibnamefont {Bonche}}, \bibinfo
  {author} {\bibfnamefont {J.}~\bibnamefont {Dobaczewski}}, \ and\ \bibinfo
  {author} {\bibfnamefont {P.-H.}\ \bibnamefont {Heenen}},\ }\href@noop {}
  {\bibfield  {journal} {\bibinfo  {journal} {Nucl. Phys.}\ }\textbf {\bibinfo
  {volume} {A542}},\ \bibinfo {pages} {355} (\bibinfo {year}
  {1992})}\BibitemShut {NoStop}%
\bibitem [{\citenamefont {D{\"o}nau}(1998)}]{donau98}%
  \BibitemOpen
  \bibfield  {author} {\bibinfo {author} {\bibfnamefont {F.}~\bibnamefont
  {D{\"o}nau}},\ }\href@noop {} {\bibfield  {journal} {\bibinfo  {journal}
  {Phys. Rev. C}\ }\textbf {\bibinfo {volume} {58}},\ \bibinfo {pages} {872}
  (\bibinfo {year} {1998})}\BibitemShut {NoStop}%
\bibitem [{\citenamefont {Almehed}\ \emph {et~al.}(2001)\citenamefont
  {Almehed}, \citenamefont {Frauendorf},\ and\ \citenamefont
  {D{\"o}nau}}]{almehed01a}%
  \BibitemOpen
  \bibfield  {author} {\bibinfo {author} {\bibfnamefont {D.}~\bibnamefont
  {Almehed}}, \bibinfo {author} {\bibfnamefont {S.}~\bibnamefont {Frauendorf}},
  \ and\ \bibinfo {author} {\bibfnamefont {F.}~\bibnamefont {D{\"o}nau}},\
  }\href@noop {} {\bibfield  {journal} {\bibinfo  {journal} {Phys. Rev.}\
  }\textbf {\bibinfo {volume} {C63}},\ \bibinfo {pages} {044311} (\bibinfo
  {year} {2001})}\BibitemShut {NoStop}%
\bibitem [{\citenamefont {Anguiano}\ \emph {et~al.}(2001)\citenamefont
  {Anguiano}, \citenamefont {Egido},\ and\ \citenamefont
  {Robledo}}]{anguiano01b}%
  \BibitemOpen
  \bibfield  {author} {\bibinfo {author} {\bibfnamefont {M.}~\bibnamefont
  {Anguiano}}, \bibinfo {author} {\bibfnamefont {J.~L.}\ \bibnamefont {Egido}},
  \ and\ \bibinfo {author} {\bibfnamefont {L.~M.}\ \bibnamefont {Robledo}},\
  }\href {\doibase 10.1016/S0375-9474(01)01219-2} {\bibfield  {journal}
  {\bibinfo  {journal} {Nucl. Phys.}\ }\textbf {\bibinfo {volume} {A696}},\
  \bibinfo {pages} {467} (\bibinfo {year} {2001})}\BibitemShut {NoStop}%
\bibitem [{\citenamefont {Dobaczewski}\ \emph {et~al.}(2007)\citenamefont
  {Dobaczewski}, \citenamefont {Stoitsov}, \citenamefont {Nazarewicz},\ and\
  \citenamefont {Reinhard}}]{dobaczewski07}%
  \BibitemOpen
  \bibfield  {author} {\bibinfo {author} {\bibfnamefont {J.}~\bibnamefont
  {Dobaczewski}}, \bibinfo {author} {\bibfnamefont {M.~V.}\ \bibnamefont
  {Stoitsov}}, \bibinfo {author} {\bibfnamefont {W.}~\bibnamefont
  {Nazarewicz}}, \ and\ \bibinfo {author} {\bibfnamefont {P.~G.}\ \bibnamefont
  {Reinhard}},\ }\href@noop {} {\bibfield  {journal} {\bibinfo  {journal}
  {Phys. Rev.}\ }\textbf {\bibinfo {volume} {C76}},\ \bibinfo {pages} {054315}
  (\bibinfo {year} {2007})}\BibitemShut {NoStop}%
\bibitem [{\citenamefont {Bender}\ \emph {et~al.}(2009)\citenamefont {Bender},
  \citenamefont {Duguet},\ and\ \citenamefont {Lacroix}}]{Bender:2008rn}%
  \BibitemOpen
  \bibfield  {author} {\bibinfo {author} {\bibfnamefont {M.}~\bibnamefont
  {Bender}}, \bibinfo {author} {\bibfnamefont {T.}~\bibnamefont {Duguet}}, \
  and\ \bibinfo {author} {\bibfnamefont {D.}~\bibnamefont {Lacroix}},\
  }\href@noop {} {\bibfield  {journal} {\bibinfo  {journal} {Phys. Rev.}\
  }\textbf {\bibinfo {volume} {C79}},\ \bibinfo {pages} {044319} (\bibinfo
  {year} {2009})}\BibitemShut {NoStop}%
\end{thebibliography}%

\end{document}